\documentclass{article}
\usepackage{arxiv}

\usepackage[utf8]{inputenc} 
\usepackage[T1]{fontenc}    
\usepackage{hyperref}       
\usepackage{url}            
\usepackage{booktabs}       
\usepackage{amsfonts}       
\usepackage{nicefrac}       
\usepackage{microtype}      
\usepackage{lipsum}

\usepackage{appendix}
\usepackage{amsmath}
\usepackage{graphicx}
\usepackage{lineno}
\usepackage{array}
\usepackage{longtable}
\usepackage{algorithm}
\usepackage{algpseudocode}
\usepackage{amssymb}
\usepackage[absolute,overlay]{textpos}

\usepackage{enumitem}
\usepackage{url}
\usepackage{adjustbox}
\usepackage[absolute,overlay]{textpos}
\usepackage{graphics}

\DeclareMathOperator*{\argmin}{argmin} 
\DeclareMathOperator*{\argmax}{argmax} 
\usepackage{mathrsfs} 

\usepackage{mathptmx}
\usepackage{stmaryrd}
\title{Estimation of Failure Probabilities via Local Subset Approximations}

\author{
  Kenan \v Sehi\'c\thanks{Corresponding author: kense@dtu.dk} \\
  Department of Applied Mathematics and Computer Science\\
  Technical University of Denmark\\
  DK-2800 Kgs. Lyngby, Denmark\\
  \And
  Mirza Karamehmedovi\'c\\
  Department of Applied Mathematics and Computer Science\\
  Technical University of Denmark\\
  DK-2800 Kgs. Lyngby, Denmark\\
}

\begin{document}
\maketitle

\begin{abstract}
We here consider the subset simulation method which approaches a failure event using a decreasing sequence of nested intermediate failure events. The method resembles importance sampling, which actively explores a probability space by conditioning the next evaluation on the previous evaluations using a Markov chain Monte Carlo (MCMC) algorithm. A Markov chain typically requires many steps to estimate the target distribution, which is impractical with expensive numerical models. Therefore, we propose to approximate each step of a Markov chain locally with Gaussian process (GP) regression. Benchmark examples of reliability analysis show that local approximations significantly improve overall efficiency of subset simulation. They reduce the number of expensive limit-state evaluations by over $80\%$. However, GP regression becomes computationally impractical with increasing dimension. Therefore, to make our use of a GP feasible, we employ the partial least squares (PLS) regression, a gradient-free reduction method, locally to explore and utilize a low-dimensional subspace within a Markov chain. Numerical experiments illustrate a significant computational gain with maintained sufficient accuracy.
\end{abstract}


\section{Introduction}
\label{S:1}
In a probabilistic framework, rare events are events with a small probability of occurrence. Accurate and efficient forecasting of rare events is essential, since an incorrect quantification can lead to a fatal failure in the modeled technological system. In general, the probability of failure $P_{\rm F}$ is defined in terms of a $d$-fold integral
\begin{equation}\label{pfd}
    P_{\rm F} = \int_{g(\theta)\leq 0} \pi (\theta) \mathrm{d}\theta,
\end{equation}
where $\theta \in \mathbb{R}^d$ is the vector of initial uncertainties for the limit-state function $g$, $\pi$ is the joint probability density function (PDF) of the input parameters $\theta$, and $g(\theta)\leq 0$ defines the failure event. We here assume the standard normal distribution for the input parameters $\theta$. By applying the Rosenblatt transformation \cite{rose} or the Nataf distribution \cite{nataf}, non-Gaussian initial uncertainties with possible correlations can be transformed to independent standard normal random variables. The limit-state function $g$ can define multiple disjoint failure regions. It describes any form of failure for a numerical or an analytical model. Typically, $g$ is a black-box model, and highly expensive to evaluate.

Reliability analysis methods based on Taylor series expansion around a design point (FORM and SORM) idealize the failure surface (the boundary of the set $\{\theta\in\mathbb{R}^d,\,\,g(\theta)\le0\}$) and do not provide an error measure \cite{Iason:2015,foam1,foam2}. A robust alternative is the simple Monte Carlo method (MC) \cite{mcbook}, which can be applied to almost any numerical model and failure surface. It approximates the probability of failure~\eqref{pfd} as the sample mean of the indicator function $\mathbb{I}(\theta)$, defined by $\mathbb{I}(\theta)=1$ if $g(\theta)\leq 0$ and $\mathbb{I}(\theta)=0$ otherwise. To accurately estimate small failure probabilities, MC requires a substantial number $N$ of limit-state evaluations. In particular, it requires $N\geq 1/(\varepsilon^2\cdot P_F)$, where $\varepsilon$ is the relative error \cite{mcbook}. For example, with $\varepsilon=0.1$, $P_F=10^{-4}$ and a one-minute numerical experiment, we would need at least around 700 days of computation. Certain variance reduction methods \cite{mcvar,mcvar1} were proposed to improve MC for low-dimensional numerical experiments.

The subset simulation method \cite{Iason:2015,zuev:2012,au:2001} is a well-known reliability approach proposed to quantify rare events for high-dimensional problems. It is also recognized as a sequential Monte Carlo because the idea is to design numerical experiments sequentially, while actively exploring the probability space. The design is related to a sequence of nested intermediate failure levels. After initial random limit-state evaluations, each new numerical experiment is designed conditioned on the samples that generated failure at the previous intermediate level. The experiment design uses a Markov chain Monte Carlo (MCMC) algorithm. Typically, for higher dimensions, the Metropolis-Hastings (MH) algorithm generates too many repeated steps that result in a substantial correlation within a Markov chain. Therefore, a modification of the MH algorithm was proposed in \cite{au:2001} that resembles the MH within Gibbs algorithm. Later, the adaptive MCMC algorithm was introduced to generate a new candidate state that is always different from the current state \cite{Iason:2015}. The adaptive MCMC algorithm assumes that the states are jointly Gaussian with a component-wise cross-correlation factor. The implementation omits the classical MH accepting criterion. The latest attempt to improve the engine of the subset simulation method was to utilize a Hamiltonian dynamic within a Markov chain \cite{hmc}. For specific benchmark cases, it demonstrated a significant gain. However, a Hamiltonian dynamic is difficult to define and solve optimally. In general, the concept is to generate multiple short Markov chains at each intermediate failure level. Because intermediate failure levels are nested, we are not required to include a burn-in period, the initial state of a Markov chain already being within the target distribution.

A Markov chain typically requires a sufficient number of states to define the target distribution adequately. In each state, the limit-state function is evaluated to test the failure criterion. This becomes impractical when the numerical evaluation of the limit state is expensive. Therefore, we approximate the limit-state function locally within the subset simulation method to efficiently quantify rare events \cite{Conrad:2016}. We assume that the limit-state function is deterministic and accessible only as a black-box model, i.e., we take the non-intrusive approach. Standard approximation methods tend either to over- or under-predict rare events and may introduce a bias if the training set is insufficient. We exploit the local regularity of the limit-state function to approximate it adequately with few samples. Global approximations within the subset simulation method include Support Vector Machines \cite{Bourinet:2011,Bourinet:2016} and Gaussian processes \cite{beck1}. The concept is mainly to train a surrogate model globally at each intermediate failure level and improve the model using a specific criterion. An alternative approach with the Multilevel Monte Carlo method (MLMC) was proposed to utilize different mesh grids at each intermediate failure level \cite{mlmc}. However, this approach does not guarantee the nestedness of intermediate failure levels, as well as requires a burn-in period.

The requirements for local approximations typically exponentially increase with the dimension, and the computation becomes infeasible. Therefore, we employ the partial least squares (PLS) regression \cite{pls2} to define a low-dimensional subspace within Markov chain steps locally. PLS maximizes the squared covariance between the low-dimensional projection of the input parameter and the limit-state value. The approach does not require gradient evaluations, which makes it suitable for expensive black-box problems. The local subset approach can be implemented easily within any MCMC algorithm.

In Section 2, we describe the subset simulation method, and we introduce local approximations based on Gaussian process regression in Section 3. Section 4 describes the implementation of the partial least squares (PLS) regression within a Gaussian process. We discuss the numerical experiments in Section 5 and offer our conclusions in Section 6.

\section{Subset Simulation Method}\label{ss}
\label{S:2}
\begin{figure}[H]
    \centering
    \includegraphics[scale=0.19]{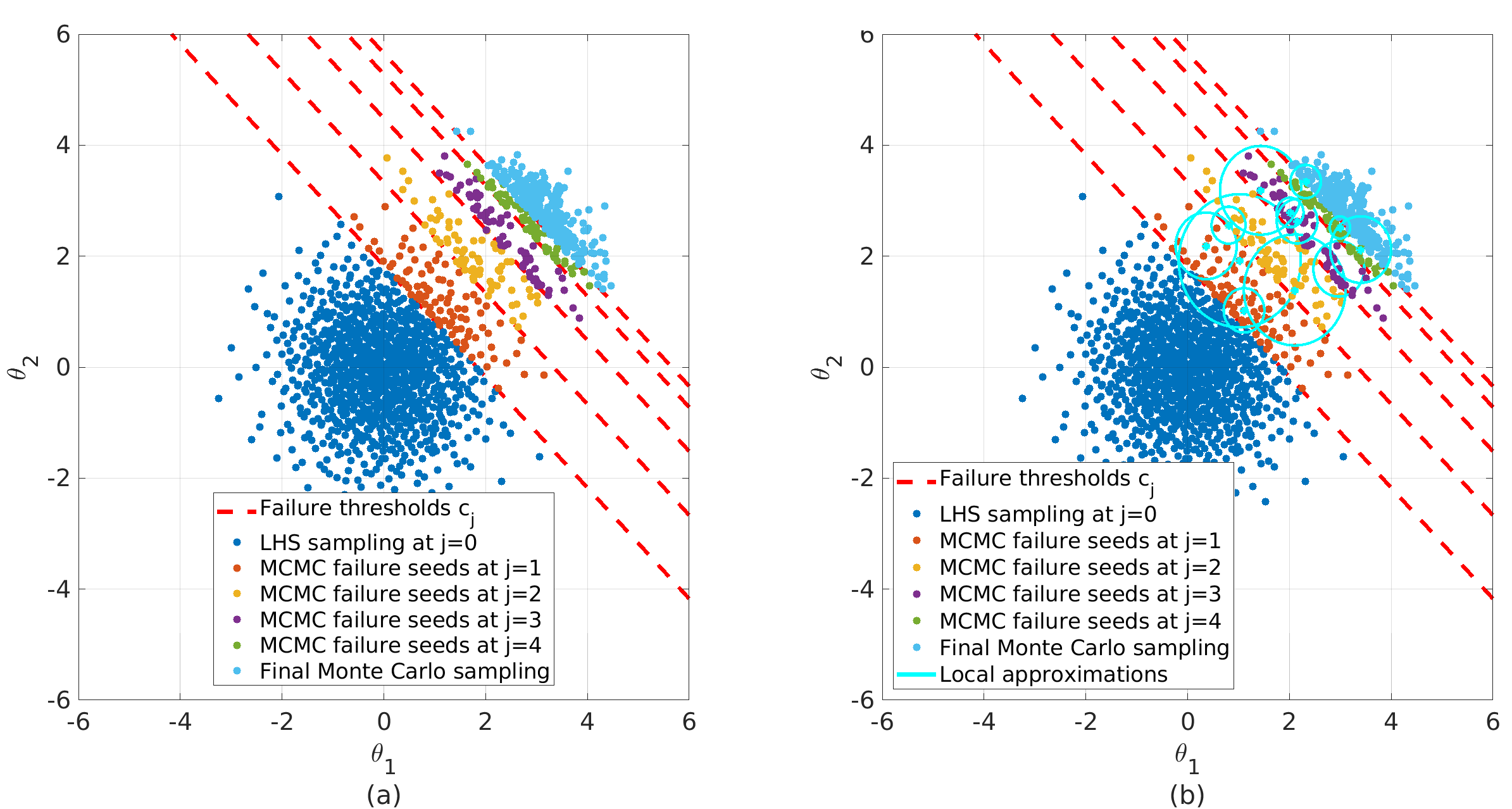}
    \caption{(a) The subset simulation method for the simple linear limit-state function. (b) Local approximations for the subset simulation method.}
    \label{slika}
\end{figure}

The failure event $F$ for the limit-state function $g$ is defined as $F=\{\theta \in \mathbb{R}^d: g(\theta)\leq 0\}$ within the probability space of the input parameters $\theta$. As illustrated in Fig.~\ref{slika}a, the idea of the Subset Simulation Method is to approach $F$ using a decreasing sequence of nested intermediate failure events, $F_1 \supset F_2 \supset~\cdots~\supset F_L = F$ for which we can write \cite{Iason:2015}
\begin{equation}
    F = \bigcap_{j=1}^{L} F_j.
\end{equation}
Hence, the probability of failure $P_{\rm F}$, Eq.~\eqref{pfd}, is estimated as a product of conditional probabilities using the intermediate failure events as
\begin{equation}
    P_{\rm F} = {\rm Pr}(F) = {\rm Pr}(\bigcap_{j=1}^{L} F_j) = \bigsqcap^L_{j=1} {\rm Pr}(F_j|F_{j-1}),
\end{equation}
where $F_0=\mathbb{R}^d$ is 'the certain event.' Initially, we generate limit-state evaluations for independent samples $\theta$ drawn from the probability density $\pi(\theta)$, and estimate the probability $P_{\rm F}^{(1)} = {\rm Pr}(F_1|F_0)$ at $j=1$ for $\theta \in F_1$. The intermediate failure probabilities $P_{\rm F}^{(j)} = \{{\rm Pr}(F_j|F_{j-1}): j = 2,~\cdots, L\}$ are then estimated by generating samples from the conditional probability distribution functions (PDFs) $\{ \pi(\theta|F_{j-1}): j = 2,~\cdots, L\}$ as \cite{Iason:2015,zuev:2012}
\begin{equation}
    \pi(\theta|F_{j-1}) = \frac{\pi(\theta)\mathbb{I}_{F_{j-1}}(\theta)}{{\rm Pr}(F_{j-1})},
\end{equation}
where $\mathbb{I}_{F_{j-1}}(\theta)$ is the indicator function for $F_{j-1}$. To generate samples from the conditional probability $\pi(\theta|F_{j-1})$, we employ a Markov chain Monte Carlo (MCMC) algorithm with the input parameters $\theta \in F_{j-1}$ as the initial state. When a Markov chain reaches its stationary state, the generated samples are identically distributed according to the conditional probability $\pi(\theta|F_{j-1})$~\cite{Iason:2015,iason46}, but not independent. As the procedure is adaptive, we set the intermediate failure probabilities to a prescribed conditional probability $p_0$, which results in the intermediate failure thresholds $c_j$. The failure events are then defined as $F_j=\{\theta \in \mathbb{R}^d: g(\theta)\leq c_j\}$, where $c_1 > c_2 > \cdots > c_L = 0$. It was demonstrated in~\cite{zuev:2012} that minimizing the coefficient of variation $\delta$ makes the prescribed conditional probability $p_0$ range between 0.1 and 0.3.

At each intermediate failure level $j$, we generate $N_s=p_0\cdot N$ Markov chains from the samples that we observe in the previous level $j-1$. Each chain generates $N/N_s-1$ steps to obtain the total of $N$ samples $\theta_j^{(1)},\dots,\theta_j^{(N)}$ from the conditional probability $\pi(\theta|F_{j-1})$. Typically, a Markov chain requires a burn-in period. However, because the initial states for Markov chains are already within the target distribution due to the nestedness of the intermediate failure events, the process is recognized as \textit{perfect simulation} and does not require a burn-in period \cite{mlmc}. 

The procedure iterates until $c_j\leq 0$, at which point the actual failure event $F=F_{L}$ is achieved. The probability of failure is then approximated using
\begin{equation}\label{glavna}
    P_{\rm F} \approx \widehat{P}_{\rm F} = p_0^{L-1}\widehat{P}_{\rm F}^{(L)},
\end{equation}
where $\widehat{P}_{\rm F}^{(L)}$ is an estimate of the final level probability $P_{\rm F}^{(L)}={\rm Pr}(F_{L}|F_{L-1})$. This estimate is obtained using the simple Monte Carlo method, as the sample mean of the indicator function $\mathbb{I}$, using the samples from $\pi(\theta|F_{L-1})$.

\subsection{Performance}
In general, the number of intermediate failure thresholds $L-1$ is random. However, for a sufficient number of samples $N$, \textbf{Lemma 1} in~\cite{cerou:2012} demonstrates that $L-1$ is actually fixed by the ratio of the logarithms
\begin{equation}\label{lemma1}
    L - 1 = \Bigg \lfloor \frac{\log \mathbb{P}( \theta \in F)}{\log p_0}\Bigg \rfloor = \Bigg \lfloor \frac{\log P_{\rm F}}{\log p_0}\Bigg \rfloor.
\end{equation}
Because the estimates of the intermediate conditional failure probabilities are correlated, the final approximation $\widehat{P}_{\rm F}$ is biased with the order of $\mathcal{O}(N^{-1})$ \cite{Iason:2015,cerou:2012}. Even for independent and identically distributed (iid) samples, it was demonstrated~\cite{cerou:2012} that bias is still present because we select intermediate failure thresholds adaptively. However, the bias seems negligible relative to the coefficient of variation of $\widehat{P}_{\rm F}$.

By using the first-order Taylor series expansion of Eq.~\eqref{glavna}~\cite{Iason:2015}, the coefficient of variation for $\widehat{P}_{\rm F}$ is estimated as
\begin{equation}
    \delta_{P_{\rm F}}^2 \approx \sum^{L}_{i=1} \sum^{L}_{j=1} \delta_n \delta_m \rho^{*}_{nm},
\end{equation}
where $\rho^*_{nm}$ is the correlation between the estimates $\widehat{P}^{(n)}_{\rm F}$ and $\widehat{P}^{(m)}_{\rm F}$. The coefficients of variation $\delta_j$ of the conditional probabilities $\widehat{P}_{\rm F}^{(j)}$ are defined by \cite{Iason:2015}:
\begin{equation}
    \delta_j = \sqrt{\frac{1-P^{(j)}_{\rm F}}{N\cdot P^{(j)}_{\rm F}}(1+\gamma_j)}.
\end{equation}
Here, $\gamma_j$ defines the auto-correlation of Markov chain states. It is estimated with the indicator function $\mathbb{I}_{\rm F_j}$ for the failure level $j$ using limit-state evaluations within a Markov chain. At the failure level $j=1$, the coefficient of variation $\delta_1$ for $\widehat{P}_{\rm F}^{(1)}$ is estimated with $\gamma_j=0$ because we use only the simple Monte Carlo evaluations.

We summarize the standard implementation of the subset simulation method in \textbf{Algorithm \ref{MCMCSS}}. The algorithm uses the adaptive MCMC implementation (\textbf{Algorithm \ref{MCMCSS0}}) that generates a candidate state $v$ that is always different from its current state $\theta_j^{(s)}$ with the assumption that $v$ and $\theta_j^{(s)}$ are jointly Gaussian with a component-wise cross-correlation factor $\rho_{d_i}$. We note the Markov chain step with $s$. The relation between the cross-correlation $\rho_{d_i}$ and the variance $\sigma_{d_i}^2$ is $\rho_{d_i} = \sqrt{1-\sigma_{d_i}^2}$. A low $\rho_{d_i}$ and a large variance $\sigma_{d_i}^2$ result in many rejected candidates, while a small variance and a large $\rho_{d_i}$ result in a high correlation between states. The cross-correlation factor $\rho_{d_i}$ is updated iteratively to keep the acceptance rate close to $0.44$, which was observed to be an optimal rate for the subset simulation method \cite{Iason:2015,zuev:2012}. The acceptance rate and the standard deviation of the proposal distribution of each component $\sigma_{d_i}$ are combined with the scaling parameter $\lambda_{\rm iter}$. This parameter is updated iteratively within Markov chain steps by using the measured accepting rate of a chain. Therefore, we skip the classical MH accepting step and focus only on the failure condition of a candidate state $v$. Given an arbitrary state $\theta_j^{(s)}$, a candidate state $v$ is generated from the multivariate normal distribution with the mean $\rho_{d_i}\theta_{j,d_i}^{(s)}$ and the standard deviation $\sqrt{1 - \rho_{d_i}^2}$. For more details, the reader should consult~\cite{Iason:2015}. It is crucial to note that the standard implementation of the subset simulation method, \textbf{Algorithm \ref{MCMCSS}}, assumes that the shape of the intermediate failure domain $F_j$ approaches continuously, with increasing index $j$, the shape of the original failure domain $F$. If this is not the case, the algorithm could end up sampling in the wrong direction \cite{susprob}.

\begin{algorithm}[H]
\footnotesize
\caption{Subset simulation method \cite{Iason:2015,au:2001}}
\label{MCMCSS}
\begin{algorithmic}[1]
\footnotesize
\Procedure{SuS}{$N$ (the number of samples in each intermediate step), $p_0 \in [0,1]$ (the conditional probability), $g$ (the limit-state function)}
    \State Generate $N$ iid samples $\theta_{j=0, i} \in \mathbb{R}^{N\times d}$ from a joint probability density function $\pi(\theta)$.
    \State Sort the samples $\theta_{j=0, i}$ in ascending order by their equivalent magnitudes of the limit-state values $g(\theta_{j=0, i})\in \mathbb{R}^{N}$.
    \State For $j=1$, define $c_1$ as the $p_0$-quantile of the evaluations $g(\theta_{j=0, i})$ and the intermediate failure threshold $F_1=\{\theta \in \mathbb{R}^d: g(\theta) \leq c_1\}$.
    \Repeat
    \State Generate $N$ samples from $\pi(\theta|F_j)$ by using $p_0\cdot N$ multiple short Markov chains for the samples that satisfy $\theta_{j-1, i} \in F_j$.
    \State Define $F_{j+1}=\{\theta \in \mathbb{R}^d: g(\theta) \leq c_{j+1}\}$ with $c_{j+1}$ as the $p_0$-quantile for $N$ generated samples by the Markov chains.
    \State j=j+1
    \Until{$c_j>0$}
    \State Identify $N_{\rm F}$ failure samples at the final level for which $\theta_{j-1, i} \in F$.
    \State Estimate the failure probability as $\widehat{P}_{\rm F} = p_0^{j-1}\frac{N_{\rm F}}{N}$.
\EndProcedure
\end{algorithmic}
\end{algorithm}

\begin{algorithm}[H]
\footnotesize
\caption{Conditional sampling from $\pi(\theta|F_j)$ \cite{Iason:2015}}
\label{MCMCSS0}
\begin{algorithmic}[1]
\footnotesize
\Procedure{Adaptive-MCMC}{$\theta_j^{(s)}$ (the previous state within a Markov chain), $\rho_j$ (the correlation), $F_j$ (the intermediate failure), $s$ (a Markov chain step)}
    \State Generate a candidate state $v \in \mathbb{R}^d$ from $\pi(\theta|F_j)$. For each coordinate $d_i$ of a candidate state $v$, generate $v_{d_i}$ from the normal distribution with the mean $\rho_{d_i}\theta_{j, d_i}^{(s)}$ and the standard deviation $\sqrt{1 - \rho_{d_i}^2}$.
    \If{$v \in F_j$}
        \State $\theta_j^{(s+1)} = v$
    \Else
        \State $\theta_j^{(s+1)} = \theta_j^{(s)}$
    \EndIf
\EndProcedure
\end{algorithmic}
\end{algorithm}

The subset simulation method with the adaptive MCMC approach, \textbf{Algorithm \ref{MCMCSS}}, requires the total number $N_{\rm Total}$ of limit-state evaluations
\begin{equation}\label{ntot}
    N_{\rm Total} = N + N \cdot (1-p_0)\cdot(L-1).
\end{equation}
For computationally demanding numerical experiments, this requirement is infeasible.  Also, $N_{\rm Total}$ increases linearly with $L$. Because the approach uses an MCMC algorithm, we need to employ multiple different runs of \textbf{Algorithm \ref{MCMCSS}} to quantify the variability of the solution, which is an additional cost.

\section{Local Subset Approximations}
\label{S:3}
Here, we propose a different approach to improve the subset simulation method. For each Markov chain proposal, we choose to use only a subset of $N_0$ nearby samples from the $N$ available samples to predict the limit-state function, see Fig.~\ref{slika}b. We describe the procedure in more detail in \textbf{Algorithm \ref{glavnialgorithm}}. We can split the algorithm into three parts. The first part, which is discussed in this section, covers how the limit-state function is locally approximated, while the second part, lines 4--15, deals with approximation errors. The last part, lines 16--20, checks whether a candidate state $v$ is within an intermediate failure region or not. 

\begin{algorithm}[H]
\footnotesize
\caption{Local subset approach with a Gaussian process for a Markov chain}
\label{glavnialgorithm}
\begin{algorithmic}[1]
\footnotesize
\Procedure{local-SuS}{$v$ (a candidate state), $\theta_j^{(s)}$ (a previous state), $S_{\rm T}=\{\theta_{j,i},g(\theta_{j,i})\}$ (a design set), $\theta_{j,i}: i=1,...,N_0$ (a sample set), $g$ (the limit-state function), N (the number of samples in each intermediate step), $N_0$ (the number of samples for a local approximation), $j$ (a failure level), $s$ (a Markov chain step)}
    \Repeat
    \State Compute the nominal approximations $\widehat{g}(v)$ and $\widehat{g}(\theta_j^{(s)})$ by the local subset approach within $\mathcal{B}(v,R)$ and $\mathcal{B}(\theta_j^{(s)},R)$ using $N_0$.
    \State Using Eq.~\eqref{errorloc}, estimate $\varepsilon^v$ and $\varepsilon^{\theta_j^{(s)}}$.
    \If{$ u \sim$ Uniform(0,1) $> \beta_{\rm T}$}
        \State Refine randomly at $v$ or $\theta_j^{(s)}$ for $a=c_j$.
    \ElsIf{$\varepsilon^v \geq \varepsilon^{\theta_j^{(s)}}$ AND $\varepsilon^v \geq \gamma_{\rm T}$}
        \State $S_{\rm T} \leftarrow$  Refine near $v$ using Eq.~\eqref{refine} for $a=0$.
    \ElsIf{$\varepsilon^{\theta_j^{(s)}} > \varepsilon^{v}$ AND $\varepsilon^{\theta_j^{(s)}} \geq \gamma_{\rm T}$}
        \State $S_{\rm T} \leftarrow$ Refine near $\theta_j^{(s)}$ using Eq.~\eqref{refine} for $a=0$.
    \EndIf
    \Until{True}
    \If{$v$ does not satisfy the nestedness condition}
        \State Use the limit-state function $g(v)$.
    \EndIf
    \If{$v \in F_j$}
        \State $\theta_j^{(s+1)}=v$
    \Else
        \State $\theta_j^{(s+1)}=\theta_j^{(s)}$
    \EndIf
\EndProcedure
\end{algorithmic}
\end{algorithm}

In line 3 we employ a Gaussian process (GP) regression using a Bayesian approximation, so the limit-state function $g$ must be smooth. Local Gaussian process regression was analysed in~\cite{Conrad:2016,gp3,gp2,gp1,gramacy}. GP regression approximates the limit-state function $g$ by a realization of an underlying Gaussian process \cite{bruno:2016},
\begin{equation}
    g(\theta) \approx \widehat{g}(\theta) = \beta_g^{\rm T} \cdot f_{\rm T}(\theta) + \sigma_g^2 Z(\theta,\omega_g),
\end{equation}
where $\beta_g^{\rm T} \cdot f_{\rm T}(\theta)$ is the trend and $\sigma_g^2$ is the variance of a model. Furthermore, $Z(\theta,\omega_g)$ is a stationary Gaussian process with $\omega_g\in\Omega$ being an elementary event from the probability space $(\Omega, \mathcal{F},\mathbb{P})$. A stationary Gaussian process is defined with zero-mean and unit-variance. In general, GP regression assumes a normal distribution over observations and utilizes a Bayesian approximation. To describe the correlation within a given sample set, we employ a covariance matrix $K_{nm} = \mathcal{K}(\theta_{j, n},\theta_{j, m}; \chi)$, where $\mathcal{K}$ is a predefined kernel function and $\chi$ are the hyperparameters such as the overall correlation of samples and the smoothness of $\mathcal{K}$.

As Fig.~\ref{slika}b suggests, to construct the GP regression locally, we need to specify the radius $R$ of a ball
\begin{equation}
    \mathcal{B}(v,R) := \{\theta_{j, i},\,\,\, \Vert \theta_{j, i} - v \Vert_2 \leq R \}
\end{equation}
centered on a candidate state $v$. The radius $R$ is chosen to include a fixed number of samples $N_0$. We select the number of samples as $N_0 = \sqrt{d}(d+1)(d+2)/2$ using the factorial design with additional few samples to improve the stability \cite{Conrad:2016}. However, for higher dimensions, the computation becomes impractical. In general, selecting an optimal sample size is a well-known problem in GP regression. We adopt $N_0=d+1$ for high-dimensional problems and, as explained later in this paper, we expect the error indicators to maintain an adequate performance even with suboptimal sample size. The vector of evaluations of the limit-state function at the samples in the ball $\mathcal{B}(v,R)$ is $Y = (Y_i = g(\theta_{j, i}))_{i=1,\dots,N_0}$. The parameters $\beta_g, \sigma_g^2$ are estimated by generalized least-squares \cite{bruno:2016}, while the hyperparameters $\chi$ are estimated by maximum likelihood estimation. 

Therefore, for a candidate state $v$ within a Markov chain, we predict locally the limit-state function $g(v)$ with $\mu_v(v)$ given by~\cite{bruno:2016}
\begin{equation}
    \mu_v(v) = f_{\rm T}(v)\cdot\beta_g + \rho(v)^{\rm T}K^{-1}(Y - \mathbf{F}_{\rm T}\beta_g),
\end{equation}
and define the variance (an uncertainty measure) $\sigma_v^2(v)$ as
\begin{equation}
    \sigma_v^2(v) = \sigma_g^2\Bigg( 1 - \langle f_{\rm T}(v)^{\rm T}\rho(v)^{\rm T} \rangle \begin{bmatrix}
0 & \mathbf{F}_{\rm T}^{\rm T}\\
\mathbf{F}_{\rm T} & K \end{bmatrix}^{-1} \begin{bmatrix}
f_{\rm T}(v) \\
\rho(v) \end{bmatrix} \Bigg).
\end{equation}
At the intermediate level $j$, the correlation $\rho(v)$ between the candidate state $v$ and the nearest $N_0$ samples is defined by $\rho(v):=(\mathcal{K}(v,\theta_{j, i};\chi))_{i=1,\dots,N_0}$. Also, $\mathbf{F}_{\rm T}$ is the information matrix for the regression model.

\subsection{Triggering Model Refinement}
Local approximations introduce errors in predictions. For the subset simulation method as a sequential approach, the errors can accumulate and severely affect the overall estimation of the probability of failure. Therefore, to control errors in predictions, we establish the refinement procedure, which should optimally use the limit-state function to improve predictions and the local sample set. In this section, we discuss when the refinement is required, while Section \ref{LMR} covers the implementation of the refinement. This part corresponds to lines 4--10 within \textbf{Algorithm \ref{glavnialgorithm}}. We treat both the candidate state $v$ and the previous state $\theta_j^{(s)}$ equally by choosing symmetric refinement criteria. The algorithm should behave identically when $v$ and $\theta_j^{(s)}$ are interchanged in order to not influence the reversibility of the transient kernel \cite{Conrad:2016}. Our implementation uses two criteria, the first being random: additional samples within the ball $\mathcal{B}(v,R)$ or the ball $\mathcal{B}(\theta_j^{(s)},R)$ are added with probability $\beta_{\rm T}$. This random refinement fits naturally with an MCMC algorithm. We write
\begin{equation}\label{random}
    \beta_{\rm T} = \beta_1\cdot s^{-\beta_0\cdot j^{\beta_2}},
\end{equation}
where $\beta_0$, $\beta_1$ and $\beta_2$ are arbitrary constants. For our numerical investigation, we define the parameters as $\beta_0 = 1$, $\beta_1 = 0.01$ and $\beta_2 = 2$.

The random refinement is essential to establish the theoretical convergence results \cite{Conrad:2016}. It does not have a significant impact on performance. As explained later in Section \ref{LMR}, the condition is generally used to refine further in the probability space within the ball $\mathcal{B}(v,R)$ or the ball $\mathcal{B}(\theta_j^{(s)},R)$, making a better spread of the samples, as samples tend to cluster in the subset simulation method. As is evident from~\eqref{random}, the random refinement occurs more frequently at the lower intermediate levels $j$.

The second criterion uses an uncertainty indicator to control errors in predictions. We compute the sensitivity of a local approximation $\widehat{g}$ using the $95\%$ confidence interval of GP predictions
\begin{equation}
    \widehat{g}^{\pm}(v) = \mu_v(v) \pm 1.96\cdot \sigma_v(v).
\end{equation}
It produces the scalar error indicators $\varepsilon^v$ and $\varepsilon^{\theta_j^{(s)}}$. 
\begin{equation}\label{errorloc}
            \varepsilon^v = \frac{\widehat{g}^{+}(v) - \widehat{g}^{-}(v)}{\mu_v(v)} \quad \varepsilon^{\theta_j^{(s)}} = \frac{\widehat{g}^{+}(\theta_j^{(s)}) - \widehat{g}^{-}(\theta_j^{(s)})}{\mu_{\theta}(\theta_j^{(s)})}.
\end{equation}
The refinement is triggered whenever one of the indicators exceeds a predefined threshold $\gamma_{\rm T}$. Between a candidate state $v$ and the previous state $\theta_j^{(s)}$, the algorithm prefers a sample with larger error estimation. The indicator is straightforward to estimate and explain. It is an efficient way to manage local approximation errors, and it is the primary source of refinement \cite{Conrad:2016,davis:2018}. 

\subsection{Local Model Refinement}\label{LMR}
When a refinement criterion is triggered, we perform refinement by selecting an optimal sample $\theta^*$ within $\mathcal{B}(v,R)$ for which we evaluate the limit-state function $g(\theta^*)$. The new sample and the corresponding evaluation are inserted into the design set $S_{\rm T}=\{\theta_{j, i}, g(\theta_{j, i})\}$. The concept is to improve the geometry of the sample set and reduce the prediction error. We employ the posterior distribution $\widehat{g}(v) \sim \mathcal{N} (\mu_v(v), \sigma_v^2(v))$ to include the information about the intermediate failure thresholds $c_j$ and the final threshold $c_j\leq0$. Because the failure probability is a binary classification, it is sufficient to introduce the probability of misclassification, for which we write \cite{bruno:2016}
\begin{equation}\label{pm}
    P_{\rm M} (v) \equiv \Phi \Bigg[ - \frac{|\mu(v)-c|}{\sigma(v)}\Bigg].
\end{equation}
Here $c$ is a generic failure threshold and $\Phi$ is the standard normal cumulative distribution function (CDF). The maximum value is achieved when the fraction (the $U$-function) tends to zero, i.e., when $P_{\rm M}=0.5$. A small value of $P_{\rm M}$ occurs when the prediction mean $\mu(v)$ is far from $c$ or when the prediction standard deviation $\sigma(v)$ is insignificant. Therefore, we select the sample $\theta^*$ that minimizes the $U$-function locally \cite{bruno:2016}
\begin{equation}\label{refine}
    \theta^* = \argmin_{\substack{\|\theta' - \Theta\|_2 \leq R\\\theta_{j, i} \in S_{\rm T}}} \frac{|\mu(\theta')-c|}{\sigma(\theta')}.
\end{equation}
Here, $\Theta$ is either a candidate state $v$ or the previous state $\theta_j^{(s)}$ depending on the refinement criteria.
\begin{figure}[ht]
    \centering
    \includegraphics[scale=0.4]{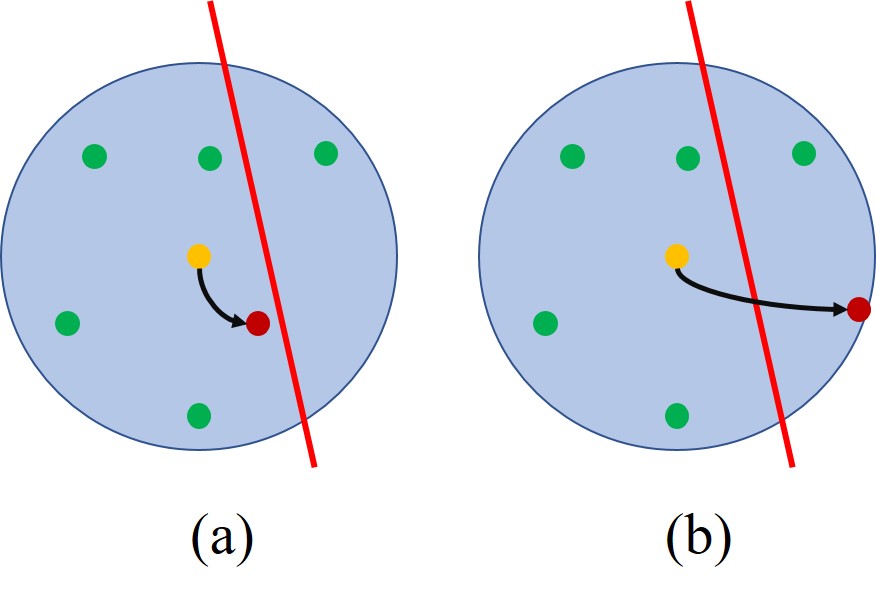}
    \caption{Illustration of the refinement using the $U$-function (the red line is a failure threshold) with (a) $a=0$, and (b) $a=-c_j$.}
    \label{refi}
\end{figure}
When the error indicator is triggered, we refine near $v$ or $\theta_j^{(s)}$ with the optimization procedure \eqref{refine} for $c=0$, see Fig.~\ref{refi}a. The constraint ensures that the new sample is within $\mathcal{B}(\Theta,R)$ to improve the current model. The inner minimization operator finds a sample that minimizes the $U$-function. As the failure is defined as $g(\theta)\leq 0$, the failure threshold $c=0$ improves the design set and the prediction globally. For the random condition, Eq.~\eqref{random}, we select $c=-c_j$ to have a sample closer to the final threshold $c_j=0$ within $\mathcal{B}(\Theta,r)$, see Fig.~\ref{refi}b.  As $-c_j \rightarrow 0$, the random refinement generates samples $\theta^*$ at the final threshold $c_j=0$.

\subsection{Failure Threshold Improvement}
The nestedness of the intermediate failure events $F_1 \supset F_2 \supset \cdots \supset F_L = F$ can be violated by local approximations. When this happens, we evaluate the limit-state function $g$ to perform a Markov chain step, line 13 of \textbf{Algorithm \ref{glavnialgorithm}}.

Recall that, for a finite number $N$ of evaluations, Eq.~\eqref{lemma1} gives the number of the intermediate failure thresholds $L-1$ as the log-ratio between the probability of failure and $p_0$. As local approximations generate errors in the intermediate failure thresholds $c_j$ and the conditional probability $p_0$, we have
\[
    \widehat{p}_0 = p_0 \pm \varepsilon.
\]
Therefore, to maintain the number of levels $L-1$ after including local approximations, the condition $p_0 \gg \varepsilon$ should be satisfied, since
\begin{equation}
    L-1 = \Bigg \lfloor \frac{\log P_{\rm F}}{\log p_0 + \log (1\pm\frac{\varepsilon}{p_0})} \Bigg \rfloor.
\end{equation}
\begin{algorithm}
\footnotesize
\caption{Intermediate failure improvement $\widehat{c}_j$}
\label{lial}
\begin{algorithmic}[1]
\footnotesize
\Procedure{Fix-$\widehat{c}_j$}{$\theta_{j,i}: i=1,\cdots,N$ - the input samples, $G: \{\widehat{g}(\theta_{j, 1}),\cdots,\widehat{g}(\theta_{j,N})\}$ - the limit-state approximation set, $g$ - the limit-state function}
    \State Set $k=0$, $N^{(k)}=0$, $G_0^{(k)}=\emptyset$ and $S^{(k)}=\emptyset$.
    \State Sort $G$ and $\theta_{j, i}$ in ascending order with $G$.
    \State Estimate the intermediate threshold $\widehat{c}_j^{(k)}$ using $p_0$ as
    \begin{align*}
        \widehat{c}_j^{(k)} = \frac{G_{Np_0}+G_{Np_0+1}}{2}.
    \end{align*}
    \Repeat
    \State Select the samples $\theta^*$ of $\theta_{j, i}$ from $N^{(k)}+1$ up to $N^{(k)} + \Delta N$ within $S^{(k)}$, for an arbitrary step $\Delta N$.
    \State Evaluate the limit-state function $g$ for $\theta\in S^{(k)}$ and collect within $G_0^{(k)}$.
    \State Update $G_{N^{(k)}+1}$ to $G_{N^{(k)} + \Delta N}$ with $G_0^{(k)}$ and sort in ascending order.
    \State Estimate the intermediate failure threshold $\widehat{c}_j^{(k+1)}$
    \begin{align*}
        \widehat{c}_j^{(k+1)} = \frac{G_{Np_0}+G_{Np_0+1}}{2}.
    \end{align*}
    \State $N^{(k+1)} = N^{(k)} + \Delta N$
    \State k=k+1
    \Until{$|\widehat{c}_j^{(k)} - \widehat{c}_j^{(k-1)}| \leq \varepsilon_s$}
\EndProcedure
\end{algorithmic}
\end{algorithm}

\begin{algorithm}
\footnotesize
\caption{Final Failure Improvement $\widehat{P}_L$ \cite{li:2010}}
\label{lial2}
\begin{algorithmic}[1]
\footnotesize
\Procedure{fix-$\widehat{P}_{\rm F}$}{$\theta_{j,i}: i=1,\cdots,N$ - the input samples, $G: \{\widehat{g}(\theta_{j, 1}),\cdots,\widehat{g}(\theta_{j, N})\}$ - the limit-state approximation set, $g$ - the limit-state function}
    \State Set $k=0$, $N^{(k)}=0$ and $S^{(k)}=\emptyset$.
    \State Estimate $\widehat{P}^{(k)}_L = \frac{N_{\rm F}}{N} = \frac{1}{N}\sum_{i=1}^{N} \mathrm{1}_{\widehat{g}(\theta)\leq 0} (\theta_{j, i})$.
    \Repeat
    \State Sort $|G|$ and $\theta_{j, i}$ in ascending order with $|G|$.
    \State Select the samples $\theta^*$ of $\theta_{j, i}$ from $N^{(k)}+1$ up to $N^{(k)} + \Delta N$ within $S^{(k)}$, for some arbitrary sample step $\Delta N$.
    \State Evaluate the limit-state function $g$ for $\theta \in S^{(k)}$.
    \State Update the failure probability
    \begin{align*}
        \widehat{P}_L^{(k)} = \widehat{P}_L^{(k-1)}+ \frac{1}{N} \sum_{\theta \in S^{(k)}} \Bigg [ -\mathbb{I}_{\widehat{g}\leq 0}(\theta) + \mathbb{I}_{g\leq 0} (\theta) \Bigg].
    \end{align*}
    \State k=k+1
    \State $N^{(k)} = N^{(k-1)} + \Delta N$
    \Until{$|\widehat{P}_L^{(k)} - \widehat{P}_L^{(k-1)}| \leq \varepsilon_s$}
\EndProcedure
\end{algorithmic}
\end{algorithm}
Additionally, we propose adaptive improvements for the intermediate failure threshold $\widehat{c}_j$ and the failure probability $\widehat{P}_F$. The intermediate failure threshold $\widehat{c}_j$ is updated after we have determined the target distribution with the adaptive MCMC approach of \textbf{Algorithm \ref{glavnialgorithm}}. In general, we iteratively replace approximations close to a failure threshold by evaluating the limit-state function for corresponding samples. Using limit-state evaluations, we update the intermediate failure threshold $\widehat{c}_j$ and the probability of failure $\widehat{P}_L$. If the procedure does not achieve a specific stopping criterion $\varepsilon_s$, it converges to the original estimations once all approximations $\widehat{g}(\theta)$ are replaced by the limit-state function $g$. See \cite{li:2010} for a rigorous convergence proof for this approach. The intermediate failure improvement is described in \textbf{Algorithm \ref{lial2}}, while for the final failure threshold we employ \textbf{Algorithm \ref{lial}} \cite{li:2010}. 

\section{Dimensionality reduction}
Typically, the computational requirements of GP regression increase with the dimension $d$, as we need a larger design set to produce an adequate result. To predict locally, we invert several times a $N_0 \times N_0$ correlation matrix, which costs $\mathcal{O}(N_0^3)$. Therefore, to increase efficiency, we employ the partial least squares (PLS) regression \cite{pls2,pls3} for line 3 of \textbf{Algorithm \ref{glavnialgorithm}}. PLS does not require gradients to explore a low-dimensional subspace, and it finds a low-dimensional projection of the input parameters $\theta$ that has significant correlation with limit-state evaluations. PLS is particularly useful when the dimension is larger than the size of the given sample set, but it requires a sufficient correlation between the input parameter $\theta$ and the limit-state evaluations \cite{pls1,pls2,pls3,pls4,pls5}. The approach combines the principal component analysis (PCA) with the ordinary least-squares regression.
\subsection{PLS background}
We define the input matrix as $\mathbf{X} \in \mathbb{R}^{N_0 \times d}$ with the corresponding evaluations of the limit-state function $Y \in \mathbb{R}^{N_0 \times 1}$. It is required to have $\mathbf{X}$ and $Y$ centered around zero, as achieved via lines 2--5 of \textbf{Algorithm \ref{pls:alg}}. The first latent component $h_1$ of the low-dimensional subspace is estimated with the optimal direction $w_1$ that maximizes the squared covariance between $h_1=\mathbf{X}w_1$ and $Y$ \cite{pls1},
\begin{equation}\label{w1}
    w_1 = \argmax_{w_1^{\rm T}w_1=1} w_1^{\rm T}\mathbf{X}^{\rm T}YY^{\rm T}\mathbf{X}w_1.
\end{equation}
The optimization problem is solved when $w_1$ is the eigenvector of the matrix $\mathbf{X}^{\rm T}YY^{\rm T}\mathbf{X}$. To obtain the second latent component $h_2$, the residual matrix $\mathbf{X}_E$ and vector $Y_{\rm F}$ are defined by subtracting from $\mathbf{X}$ and $Y$ their rank-one approximations using $t_1$  \cite{pls1}
\[
    \mathbf{X}_E = \mathbf{X} - h_1p_1^{\rm T},
\]
\[
    Y_{\rm F} = Y - b_1h_1.
\]
\begin{algorithm}[ht]
\footnotesize
\caption{PLS1 \cite{pls4}}
 \label{pls:alg}
\begin{algorithmic}[1]
\footnotesize
\Procedure{PLS1}{$\mathbf{X} \in \mathbb{R}^{N_0\times d}$ (the input matrix), $Y \in \mathbb{R}^{N_0\times 1}$ (the limit-state evaluations)}
    \State Compute the mean for $\mathbf{X}$: $\mu_X = \frac{1}{N_0}\sum_{i=1}^{N_0} \theta_{j, i}$.
    \State Compute the mean for $Y$: $\mu_Y = \frac{1}{N_0}\sum_{i=1}^{N_0} g(\theta_{j, i})$.
    \State Center $\mathbf{X}$: $\mathbf{X} = \mathbf{X} - 1\cdot \mu_X^{\rm T}$.
    \State Center $Y$: as $Y = Y - 1\cdot \mu_Y^{\rm T}$.
    \State Set $\mathbf{X}_E = \mathbf{X}$, $Y_E = Y$, $k=1$.
    \Repeat
    \State Compute the weights: $w_k = \mathbf{X}_E^{\rm T}Y_E/||\mathbf{X}_EY_E||$.
    \State Compute the score as $h_k = \mathbf{X}_Ew_k$.
    \State Compute the load as $p_k = \mathbf{X}_E^{\rm T}h_k/(h_k^{\rm T}h_k)$.
    \State Compute the regression coefficients $b_k = h_k^{\rm T}Y_E/(h_k^{\rm T}h_k)$.
    \State $\mathbf{X}_E \leftarrow \mathbf{X}_E - h_kp_k^{\rm T}$. 
    \State $Y_E \leftarrow Y_E - b_kh_k$.
    \State k = k + 1
    \Until{$||Y_E||\leq \varepsilon_y$}
\EndProcedure
\end{algorithmic}
\end{algorithm}

Here $p_1 \in \mathbb{R}^d$ is a load vector, defined in line 10 of \textbf{Algorithm \ref{pls:alg}} and $b_1$ is the corresponding regression coefficient, defined in line 11. The computation is iterative and stops when a criterion such as $\|Y_{\rm F}\| \leq \varepsilon_y$ is fulfilled \cite{pls4}. A more robust assessment can be made using cross-validation. When a stopping criterion is fulfilled, \textbf{Algorithm \ref{pls:alg}} provides the load matrix $\mathbf{P}=[p_1,...,p_r]\in  \mathbb{R}^{d \times r}$, the score matrix $\mathbf{H}=[h_1,...,h_r]\in\mathbb{R}^{N_0 \times r}$ and the weight matrix $\mathbf{W}=[w_1,...,w_r]\in\mathbb{R}^{N_0 \times r}$, where $r$ is the dimension of a low-dimensional subspace. The PLS low-dimensional subspace $\mathbb{R}^r$ is spanned by the columns of the PLS-weight matrix $\mathbf{R}_{\rm PLS}=\mathbf{W}(\mathbf{P}^{\rm T}\mathbf{W})^{-1} \in \mathbb{R}^{N_0 \times r}$. The definition of the PLS weight matrix includes the weight matrix $\mathbf{W}$, which defines the correlation between the input and the limit-state response, as well as the score matrix $\mathbf{P}$, which defines the regression relation between the input matrix $\mathbf{X}$ and the corresponding projection onto a low-dimensional subspace. Therefore, we rotate the input matrix $\mathbf{X} \in \mathbb{R}^{N_0\times d}$ by the PLS-weight matrix $\mathbf{R}_{\rm PLS}$ to discover the low-dimensional projection $\mathbf{H}_{\rm PLS}\in\mathbb{R}^{N_0\times r}$~\cite{pls5}
\begin{equation}
    \mathbf{H}_{\rm PLS} = \mathbf{X}\mathbf{R}_{\rm PLS}.
\end{equation}
Finally, instead of training a Gaussian process on the original space defined for the input matrix $\mathbf{X} \in \mathbb{R}^{N_0\times d}$, we design efficiently a Gaussian process using the low-dimensional projection $\mathbf{H}_{\rm PLS}\in\mathbb{R}^{N_0\times r}$ of the input matrix with the limit-state evaluations $Y \in \mathbb{R}^{N_0\times 1}$. This results in a smaller number of hyperparameters $\chi$ for a stationary anisotropic covariance matrix, as $r\ll d$. Because the training procedure of a Gaussian process is done on a low-dimensional subspace, we need a smaller sample size $N_0$ for a sufficient design.

\section{Examples of Application}
We evaluate the proposed local subset approach with four numerical experiments in low and high dimensions. We mainly compare the performance of the approach with the standard implementation of the subset simulation method that uses the adaptive MCMC algorithm, as explained in Section \ref{ss}. The local subset algorithm is implemented in MATLAB, and it is integrated in the algorithm of the subset simulation method provided by the Engineering Risk Analysis Group (Technical University of Munich) \cite{eralink}. Our MATLAB codes can be found at \url{https://github.com/ksehic/Local-Approximations-for-SuS}. We have there implemented both Gaussian process regression and polynomial regression. 
\begin{figure}
    \centering
    \includegraphics[scale=0.32]{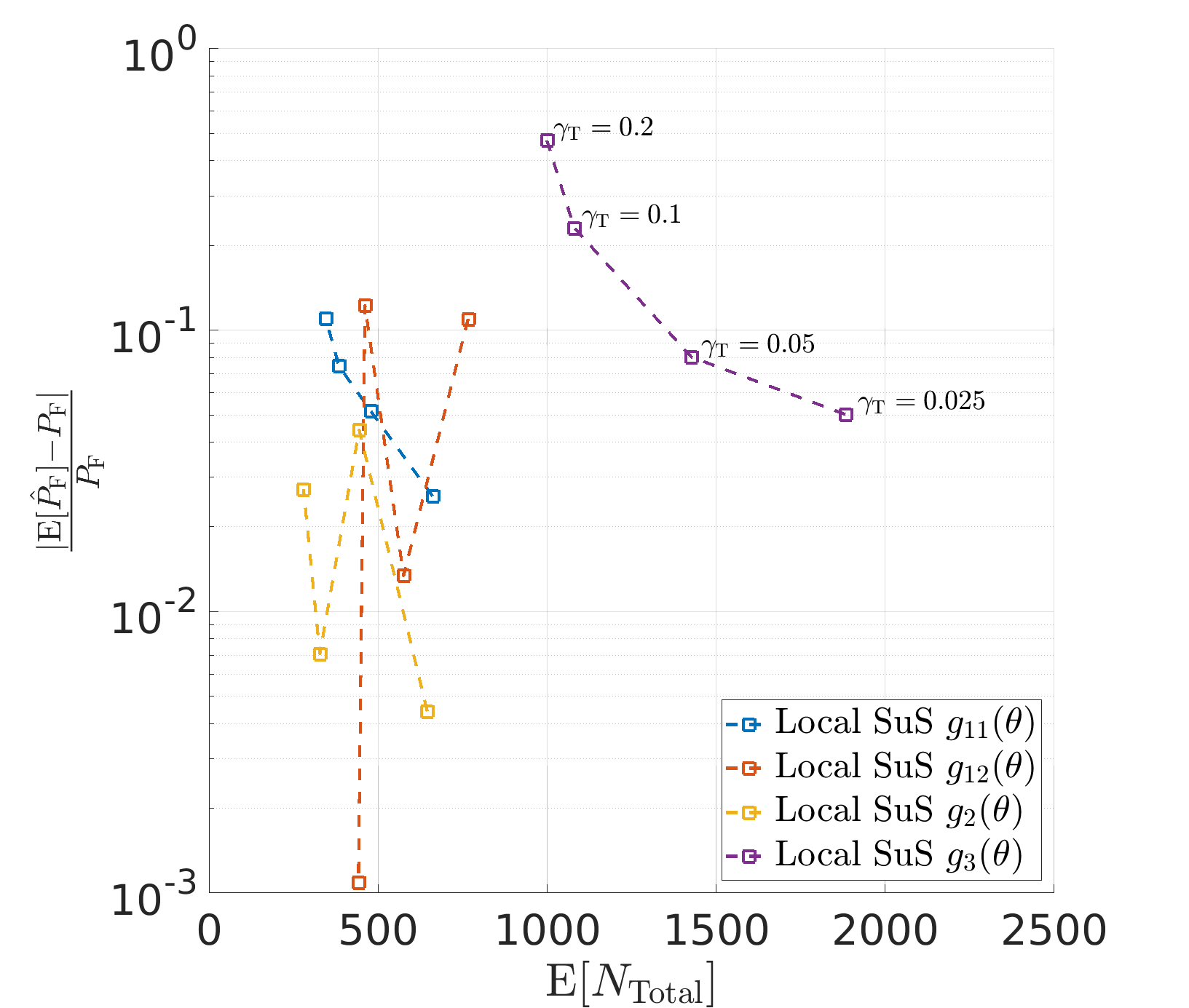}
    \caption{The relative error as a function of the average number of evaluations, over $20$ independent runs, for the two-dimensional examples 1--3 and different error thresholds $\gamma_{\rm T}$.}
    \label{gerror}
\end{figure}

To achieve an adequate initial spread of samples, we employ Latin hypercube sampling (LHS) using the built-in MATLAB function \texttt{lhsdesign}. We sample the unit hypercube $[0,1]^N$ and then map the samples to the original variable space by the inverse cumulative distribution function of the marginals. After $M$ initial limit-state evaluations, we use local approximations of LHS proposals to replace direct evaluations of the limit-state function $g$, \textbf{Algorithm \ref{start}}. We select $M$ heuristically as $M=0.1\cdot N$, which additionally improves the overall efficiency. For an LHS proposal with a substantial uncertainty in the prediction (i.e., $\gamma_{\rm T} \geq 5\%$), we employ the limit-state function $g$, lines 4--9 in \textbf{Algorithm \ref{start}}. Typically, if we increase the error threshold, the relative error increases with fewer limit-state evaluations. However, for certain numerical experiments, such as the nonlinear limit-state function from Example 1, a higher error threshold generates better results. This can be related to insufficient refinement and inadequate spread of the samples. Before sampling from the conditional probability $\pi(\theta|F_1)$ at $j=2$, the intermediate failure threshold $c_1$ is improved by \textbf{Algorithm \ref{lial}}.

\begin{algorithm}[ht]
\footnotesize
\caption{Local approximations after $k_0=M$ initial limit-state evaluations}
\label{start}
\begin{algorithmic}[1]
\footnotesize
\Procedure{local-start}{$v_0$ (an LHS proposal), $S_k=\{\theta_{j,i},g(\theta_{j,i})\}$ (a design set), $\theta_{j,i}: i=1,...,N_0$ (a sample set), $g$ (the limit-state function), N (the number of samples in each intermediate failure step), $N_0$ (the number of samples for a local approximation)}
    \State Estimate the nominal approximation $\widehat{g}(v_0)$ by a Gaussian process locally within $\mathcal{B}(v_0,R)$ using $N_0$.
    \State Compute the error indicator $\varepsilon^v$ by Eq.~\eqref{errorloc}.
    \If{$\varepsilon^v < \gamma_{\rm T}$}
        \State $S_{k+1} \leftarrow \{v_0,\widehat{g}^+(v_0)\}$.
    \Else
        \State Evaluate the limit-state function $g(v)$.
        \State $S_{k+1} \leftarrow \{v_0,g^+(v_0)\}$.
    \EndIf
    \If{$k_0==N$}
        \State j=j+1
        \State Define $\widehat{c}_j$ as the $p_0$-quantile of the evaluation part of $S_k$.
        \State Improve $\widehat{c}_j$ by \textbf{Algorithm \ref{lial}}.
    \EndIf
\EndProcedure
\end{algorithmic}
\end{algorithm}

The results are computed from 20 independent simulation runs. We fix the seed numbers to fairly compare the performances of the local subset approach and the standard implementation. In all examples, we select the typical values $N=1000$ and $p_0=0.1$~\cite{zuev:2012}. However, in certain situations, we select different values to investigate their contributions in the estimations. For Gaussian process regression, we use the constant trend with the anisotropic squared exponential kernel. Note that the acceptable relative error in reliability analysis can range up to $30\%$ due to typically small values of the probability of failure. Our probability of failure is nominal because we do not include all possible uncertainties. The presence of safety coefficients is inevitable in realistic structural design. 

\subsection{Example 1 - Simple limit-state function}
Here, we consider the limit-state surface defined by a linear function \cite{hmc,Iason:2015}
\begin{equation}\label{lin1}
    g_{11}(\theta) = 4 - \frac{1}{\sqrt{d}}\sum_{n=1}^d \theta_n,
\end{equation}
and its non-linear version \cite{hmc}
\begin{equation}\label{nonlin1}
    g_{12}(\theta) = 4 - \frac{\kappa}{4}(\theta_1 - \theta_2)^2 - \frac{1}{\sqrt{d}}\sum_{n=1}^d \theta_n.
\end{equation}
The parameter $\kappa$ controls the non-linearity of the function. We estimate the failure probabilities exactly as $P_{\rm F}^{g_{11}} = 3.17 \times 10^{-5}$ and $P_{\rm F}^{g_{12}} = 6.41 \times 10^{-5}$ for $\kappa = 0.2$ \cite{hmc}. In this example, we are able to explore the performance of the local subset approach in varying dimensions because the final failure estimation does not depend on the dimension $d$. 

\begin{figure}
    \centering
    \includegraphics[scale=0.3]{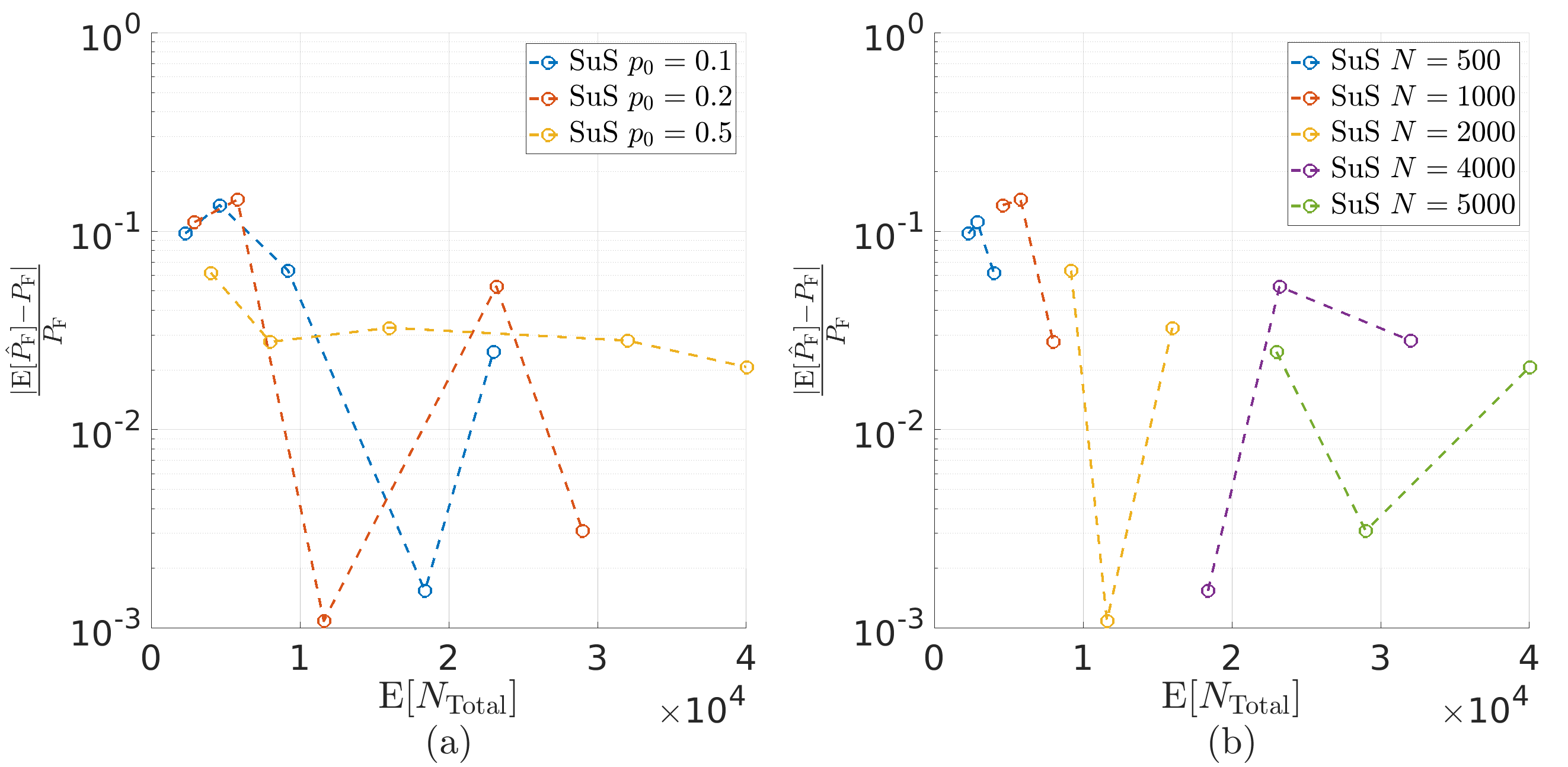}
    \caption{For the linear limit-state function, Eq.~\eqref{lin1}, the relative error of the standard implementation as a function of the average number of evaluations, over $20$ independent runs, for (a) the conditional probabilities $p_0 = (0.1, 0.2, 0.5)$ with the corresponding number of samples $N = (500, 1000, 2000, 4000, 5000)$, and (b) the number of samples $N = (500, 1000, 2000, 4000, 5000)$ with the corresponding conditional probabilities $p_0 = (0.1, 0.2, 0.5)$.}
    \label{g6sus}
\end{figure}

\begin{figure}
    \centering
    \includegraphics[scale=0.3]{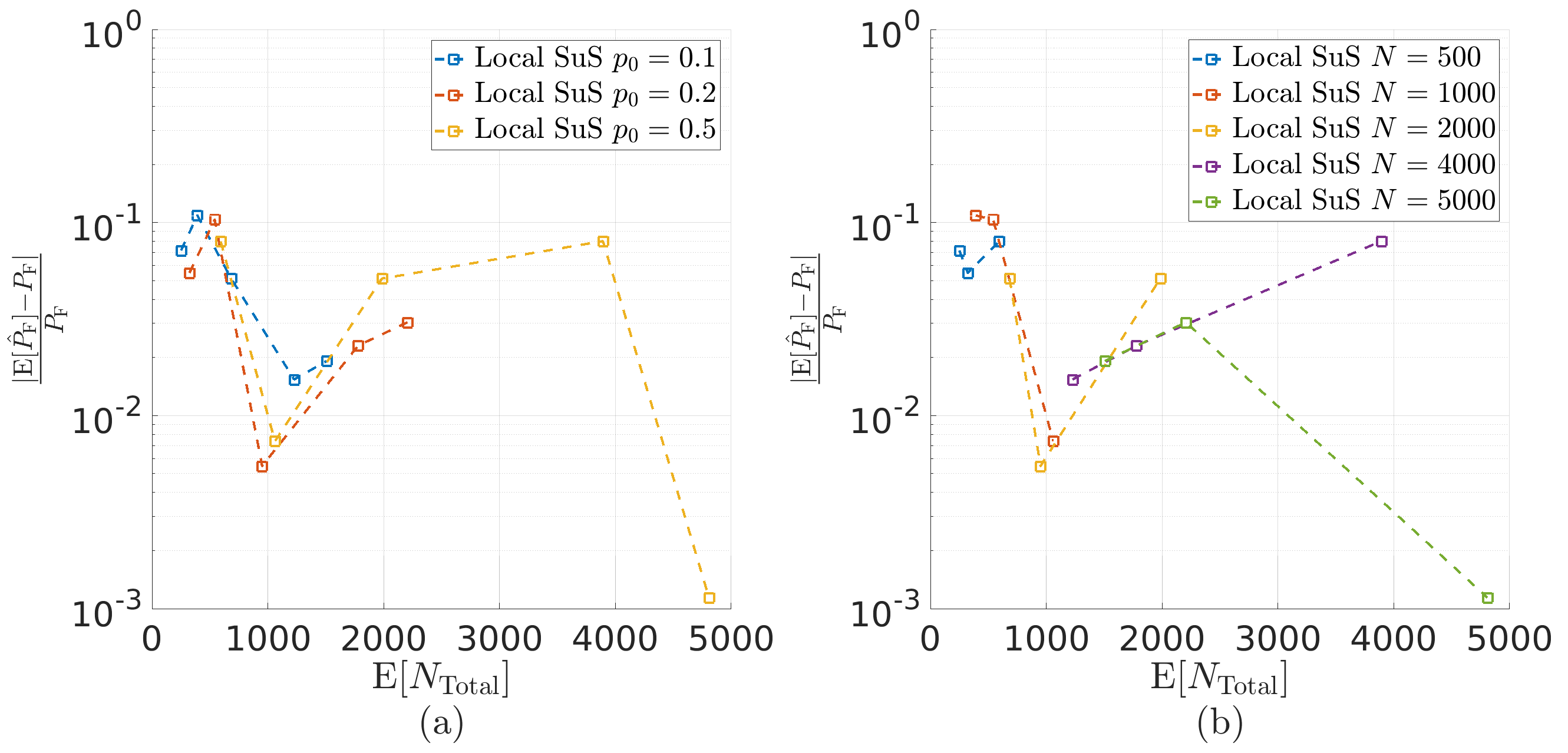}
    \caption{For the linear limit-state function, Eq.~\eqref{lin1}, the relative error of the local subset approach as a function of the average number of evaluations, over $20$ independent runs, for (a) the conditional probabilities $p_0 = (0.1, 0.2, 0.5)$ with the corresponding number of samples $N = (500, 1000, 2000, 4000, 5000)$, and (b) the number of samples $N = (500, 1000, 2000, 4000, 5000)$ with the corresponding conditional probabilities $p_0 = (0.1, 0.2, 0.5)$.}
    \label{g6loc}
\end{figure}

\begin{figure}
    \centering
    \includegraphics[scale=0.3]{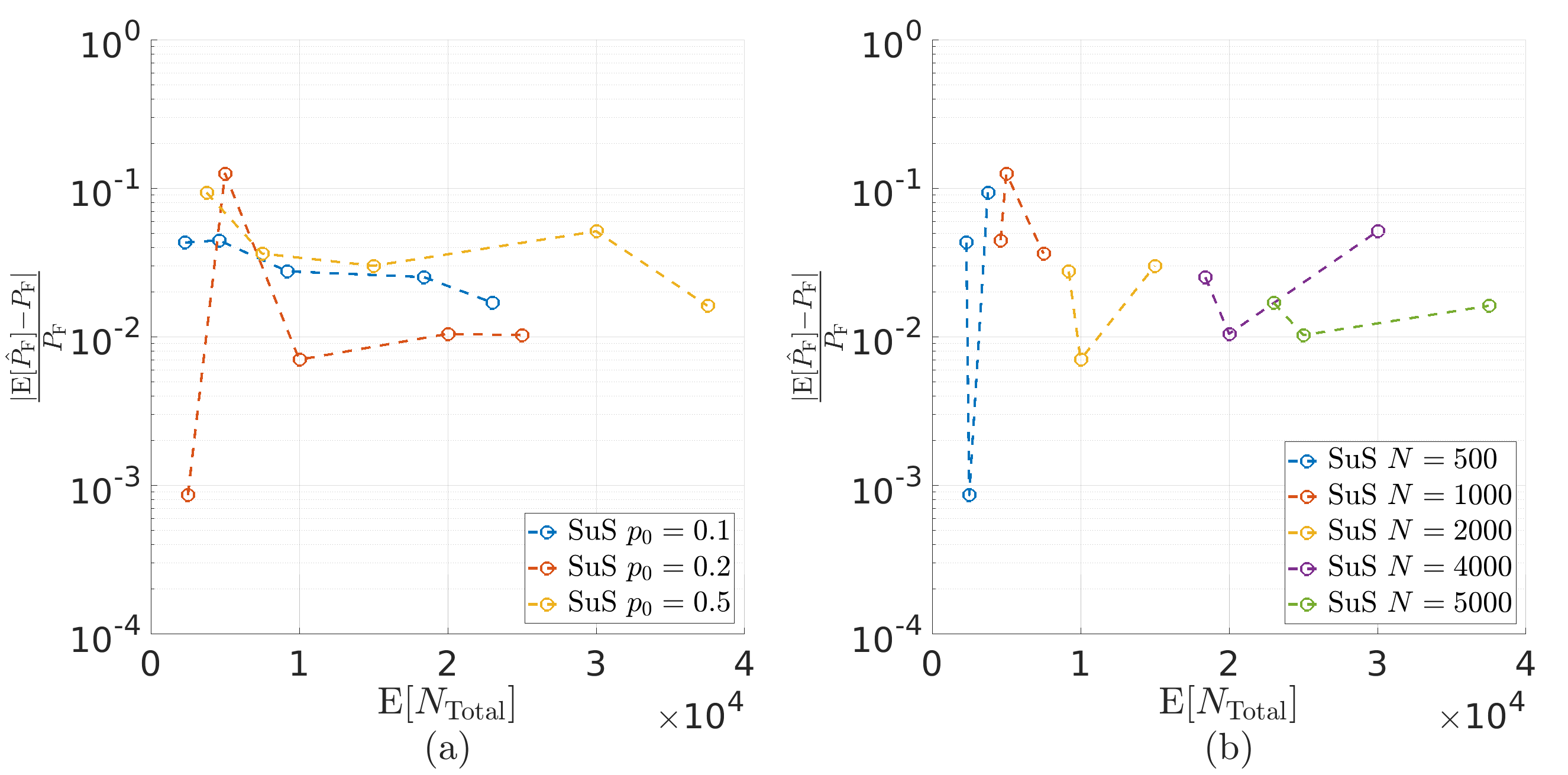}
    \caption{For nonlinear limit-state function, Eq.~\eqref{nonlin1}, the relative error of the standard implementation as a function of the average number of evaluations, over $20$ independent runs, for (a) the conditional probabilities $p_0 = (0.1, 0.2, 0.5)$ with the corresponding number of samples $N = (500, 1000, 2000, 4000, 5000)$, and (b) the number of samples $N = (500, 1000, 2000, 4000, 5000)$ with the corresponding conditional probabilities $p_0 = (0.1, 0.2, 0.5)$.}
    \label{g8sus}
\end{figure}

\begin{figure}
    \centering
    \includegraphics[scale=0.3]{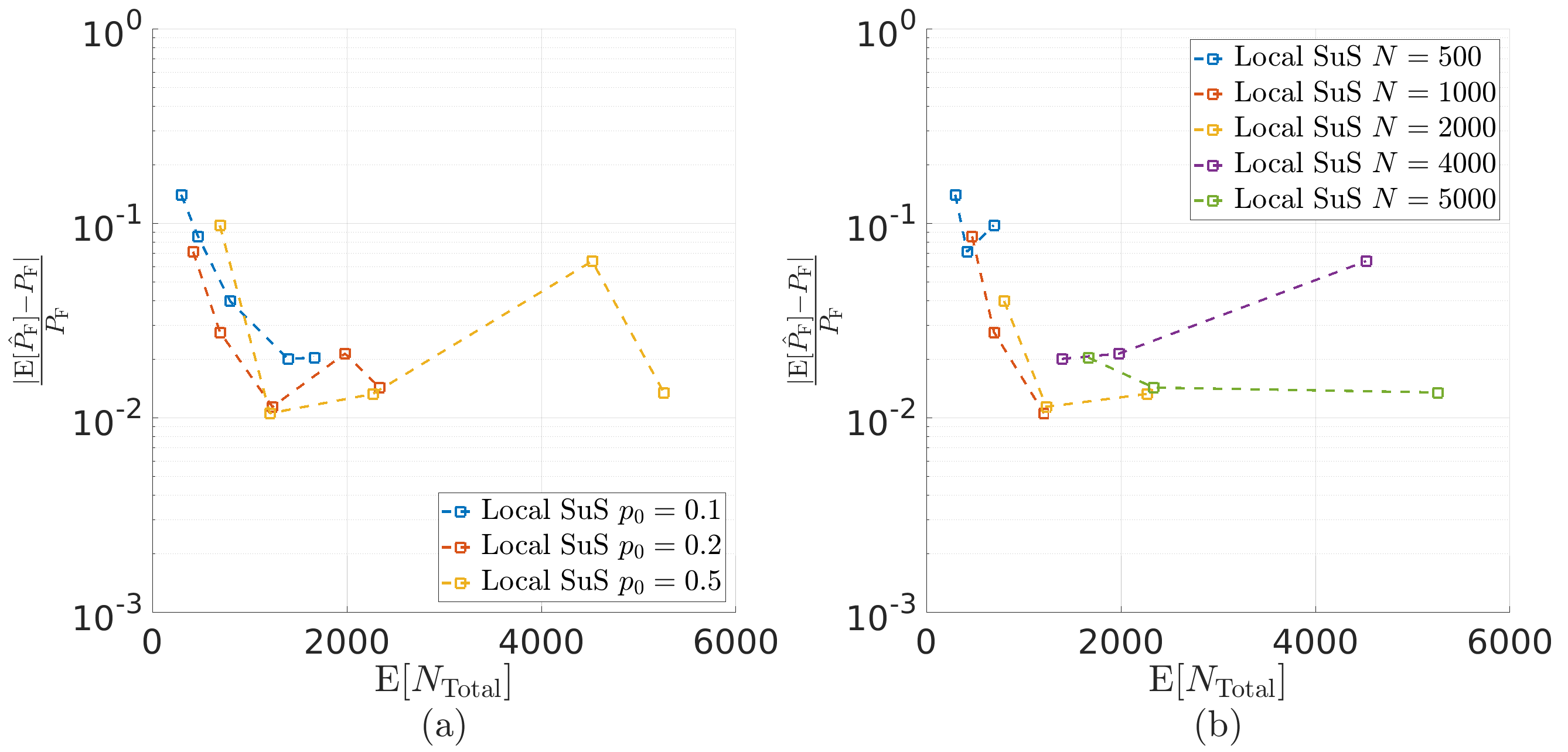}
    \caption{For the nonlinear limit-state function, Eq.~\eqref{nonlin1}, the relative error of the local subset approach as a function of the average number of evaluations, over $20$ independent runs, for (a) the conditional probabilities $p_0 = (0.1, 0.2, 0.5)$ with the corresponding number of samples $N = (500, 1000, 2000, 4000, 5000)$, and (b) the number of samples $N = (500, 1000, 2000, 4000, 5000)$ with the corresponding conditional probabilities $p_0 = (0.1, 0.2, 0.5)$.}
    \label{g8loc}
\end{figure}

The local subset approach reduces the total number of evaluations by over $89\%$ on average for the low-dimensional numerical experiments, while keeping the relative error $\varepsilon_0$ with respect to the standard implementation at less than $3\%$, see Table \ref{table:1}. In comparison with the exact solutions, the relative errors are up to $11\%$. For the standard implementation, Figs.~\ref{g6sus} and \ref{g8sus} illustrate the relation between the relative error and the average number of limit-state evaluations for different values of $p_0$ and $N$. Figures \ref{g6loc} and \ref{g8loc} show the same for the local subset approach. The local subset approach outperforms the standard implementation with significantly fewer limit-state evaluations. In general, for the local subset approach, we note that the conditional probability $p_0=0.5$ and the initial number of limit-state evaluations $N=5000$ achieve the minimal relative error for the linear limit-state function, while for the nonlinear version we have the best values $p_0=0.2$ and $N=2000$. 

In higher dimension,~e.g., $d=10$, the relative error with respect to the standard implementation increases. However, as the standard implementation overpredicts the exact solution, the relative error with respect to the exact solution decreases, see Table \ref{table:12}. The efficiency is above $82\%$ with the relative error less than $9\%$. However, the computational demands for predictions become intensive. As previously explained, this is the main reason to include PLS in predictions. (Nevertheless, as the results show, local approximations with Gaussian process regression can be used in higher dimensions, however with longer calculations.)
\begin{table}
\normalsize
\begin{adjustbox}{width=\columnwidth,center}
\begin{tabular}{c c c c c c c c c c} 
\hline\hline 
Case & $\mathbb{E}[P_{\rm F}^{\rm MC}]$ & $\mathbb{E}[P_{\rm F}^{\rm SuS}]$ & $\mathbb{E}[P_{\rm F}^{\rm Local}]$ & $\sigma[P_{\rm F}^{\rm Local}]$ & $\varepsilon$ & $\varepsilon_0$ & $\mathbb{E}[\widehat{N}_{0}]$ & $\mathbb{E}[\widehat{N}_{\rm Total}$] & $\mathbb{E}[N_{\rm Total}$]\\ [0.5ex] 
\hline 
$g_{11}(\theta)$ for $d=2$& 3.2e-5 & 3.6e-5 & 3.5e-5 & 1.2e-5 & 0.11 & 0.03  &196 & 391.6 & 4600\\ 
$g_{12}(\theta)$ for $d=2$& 6.4e-5 & 6.7e-5 & 6.9e-5 &  2.3e-5 & 0.08 & 0.03 & 190 & 469.7 & 4600\\ [1ex] 
\hline 
\end{tabular}
\end{adjustbox}
\caption{Local subset approach with a Gaussian process for Eqs.~\eqref{lin1} and \eqref{nonlin1} averaged over $20$ independent runs for $p_0=0.1$ and $N=1000$.} 
\label{table:1} 
\end{table}

\begin{table}
\normalsize
\begin{adjustbox}{width=\columnwidth,center}
\begin{tabular}{c c c c c c c} 
\hline\hline 
Case & $\mathbb{E}[P_{\rm F}^{\rm SuS}]$ & $\mathbb{E}[P_{\rm F}^{\rm Local}]$ & $\sigma[P_{\rm F}^{\rm Local}]$ & $\varepsilon_0$ & $\mathbb{E}[\widehat{N}_{\rm Total}$] & $\mathbb{E}[N_{\rm Total}$]\\ [0.5ex] 
\hline 
$g_{11}(\theta)$ for $d=2$ & 3.6e-5 & 3.5e-5 & 1.2e-5 & 0.03 & 391.6 & 4600\\ 
$g_{11}(\theta)$ for $d=5$  & 3.6e-5 & 3.3e-5 & 1.0e-5 & 0.09 &  503.3 & 4600\\ 
$g_{11}(\theta)$ for $d=10$ & 3.6e-5 & 3.3e-5 & 1.0e-5 & 0.09  &  805.2 & 4600\\ [1ex]
\hline 
\end{tabular}
\end{adjustbox}
\caption{Local subset approach with a Gaussian process for Eq.~\eqref{lin1} in higher dimensions averaged over $20$ independent runs for $p_0=0.1$ and $N=1000$.} 
\label{table:12} 
\end{table}

Therefore, we use the partial least squares (PLS) regression. Initially, we employ \textbf{Algorithm \ref{pls:alg}} with all samples in the set to estimate a low-dimensional subspace globally. We project the samples onto the global low-dimensional subspace and select the nearest samples to a projected candidate state $v$. To define a local low-dimensional subspace, the nearest samples and the corresponding limit-state evaluations are processed by \textbf{Algorithm \ref{pls:alg}}. The local low-dimensional subspace is used to train a Gaussian process efficiently.

Table \ref{table:13} shows the results for the linear and nonlinear limit-state functions of Example 1, with $d=100$. The efficiency for higher dimensions drops to $34\%$ with the relative error less than $90\%$. The relative error is substantial, but the failure level of $10^{-5}$ is accurately estimated. For demanding computations, the improvement of $34\%$ can make a significant difference. By increasing $p_0 = 0.5$, we reduce the relative error to less than $22\%$, but with the efficiency of $26.8\%$ and $31.8\%$ respectively. If we include more points in the local regression, the efficiency increases to over $50\%$, but the relative errors increase to around $40\%$. The results for $N=5000$ are significantly better, with the overall efficiency above $57.2\%$ and with relative errors at $44\%$ and $53\%$, respectively.

\begin{table}
\normalsize
\begin{adjustbox}{width=\columnwidth,center}
\begin{tabular}{c c c c c c c} 
\hline\hline 
Case & $\mathbb{E}[P_{\rm F}^{\rm SuS}]$ & $\mathbb{E}[P_{\rm F}^{\rm Local}]$ & $\sigma[P_{\rm F}^{\rm Local}]$ & $\varepsilon_0$ & $\mathbb{E}[\widehat{N}_{\rm Total}]$ & $\mathbb{E}[N_{\rm Total}$]\\ [0.5ex] 
\hline 
(a) $g_{11}(\theta)$ & 3.0e-5 & 0.2e-5 &  1.6e-6 & 0.9 & 3022.7& 4600\\ 
(a) $g_{12}(\theta)$ & 5.9e-5 & 0.6e-5 & 4.0e-6 & 0.9 & 2818.4 & 4600\\ 
\hline
(b) $g_{11}(\theta)$ & 3.0e-5 & 2.3e-5 & 7.6e-6 & 0.2 & 5856.4 & 8000\\ 
(b) $g_{12}(\theta)$ & 5.9e-5 & 4.5e-5 & 1.4e-5 & 0.2 & 5451.7 & 8000\\ 
\hline 
(c) $g_{11}(\theta)$ & 2.8e-5 & 1.6e-5 & 2.8e-6 & 0.4 & 9838.2 & 2.3e3\\ 
(c) $g_{12}(\theta)$ & 5.7e-5 & 2.7e-5 & 4.5e-6 & 0.5 & 9697.2 & 2.3e3\\ 
\hline 
\end{tabular}
\end{adjustbox}
\caption{Local PLS-Gaussian process approximations for Eq.~\eqref{lin1} and Eq.~\eqref{nonlin1} with $d=100$ for (a) $p_0=0.1$ and $N=1000$, (b) $p_0=0.5$ and $N=1000$, and (c) $p_0=0.1$ and $N=5000$.} 
\label{table:13} 
\end{table}

\begin{figure}[ht]
    \centering
    \includegraphics[scale=0.19]{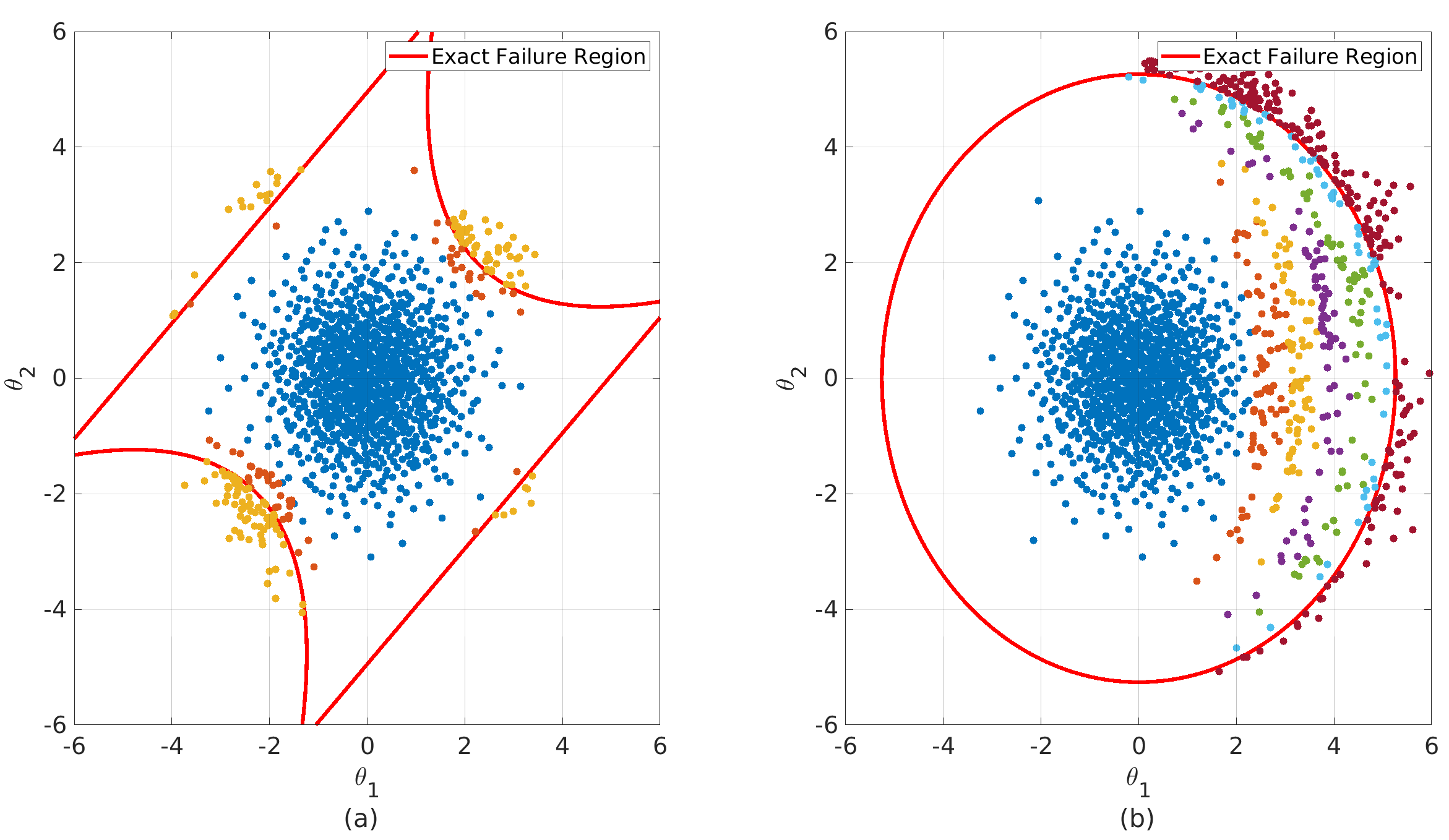}
    \caption{The subset simulation method with the local subset approach for (a) four failure branches function, and (b) hypersphere limit-state function.}
    \label{2g}
\end{figure}

\subsection{Example 2 - Four failure branches function}
A system with four distinct component limit-states \cite{bruno:2016} is a common benchmark in reliability analysis, and we can describe it with the normal distribution as 
\begin{equation}\label{ffbf}
    g_2(\theta) = \min\begin{cases}
             3 + 0.1(\theta_1-\theta_2)^2 - \frac{\theta_1+\theta_2}{\sqrt{2}},  \\
             3 + 0.1(\theta_1-\theta_2)^2 + \frac{\theta_1+\theta_2}{\sqrt{2}},  \\
             \theta_1 - \theta_2 + \frac{7}{\sqrt{2}}, \\
             \theta_2 - \theta_1 + \frac{7}{\sqrt{2}}. 
       \end{cases}
\end{equation}
\begin{table}
\normalsize
\begin{adjustbox}{width=\columnwidth,center}
\begin{tabular}{c c c c c c c c c c} 
\hline\hline 
Case & $\mathbb{E}[P_{\rm F}^{\rm MC}]$ & $\mathbb{E}[P_{\rm F}^{\rm SuS}]$ & $\mathbb{E}[P_{\rm F}^{\rm Local}]$ & $\sigma[P_{\rm F}^{\rm Local}]$ & $\varepsilon_0$ & $\varepsilon$ & $\mathbb{E}[\widehat{N}_{0}]$ & $\mathbb{E}[\widehat{N}_{\rm Total}]$ & $\mathbb{E}[N_{\rm Total}]$\\ [0.5ex] 
\hline 
$g_{2}(\theta)$ & 2.3e-3 & 2.4e-3 & 2.3e-3 & 4.0e-4 & 0.02 & 0.04 & 222 & 348.6 & 2800\\[1ex] 
\hline 
\end{tabular}
\end{adjustbox}
\caption{Local polynomial approximation results for the four failure branches function averaged over 20 independent runs with $p_0=0.1$ and $N=10000$.} 
\label{table:2} 
\end{table}
Two components are linear, while the remaining two components are described with the parabolic shapes. The reference probability of failure is estimated to $P_{\rm F}^{\rm MC} = 2.26 \times 10^{-3}$ using the simple Monte Carlo method with $N_{\rm MC} = 1 \times 10^6$.

\begin{figure}
    \centering
    \includegraphics[scale=0.3]{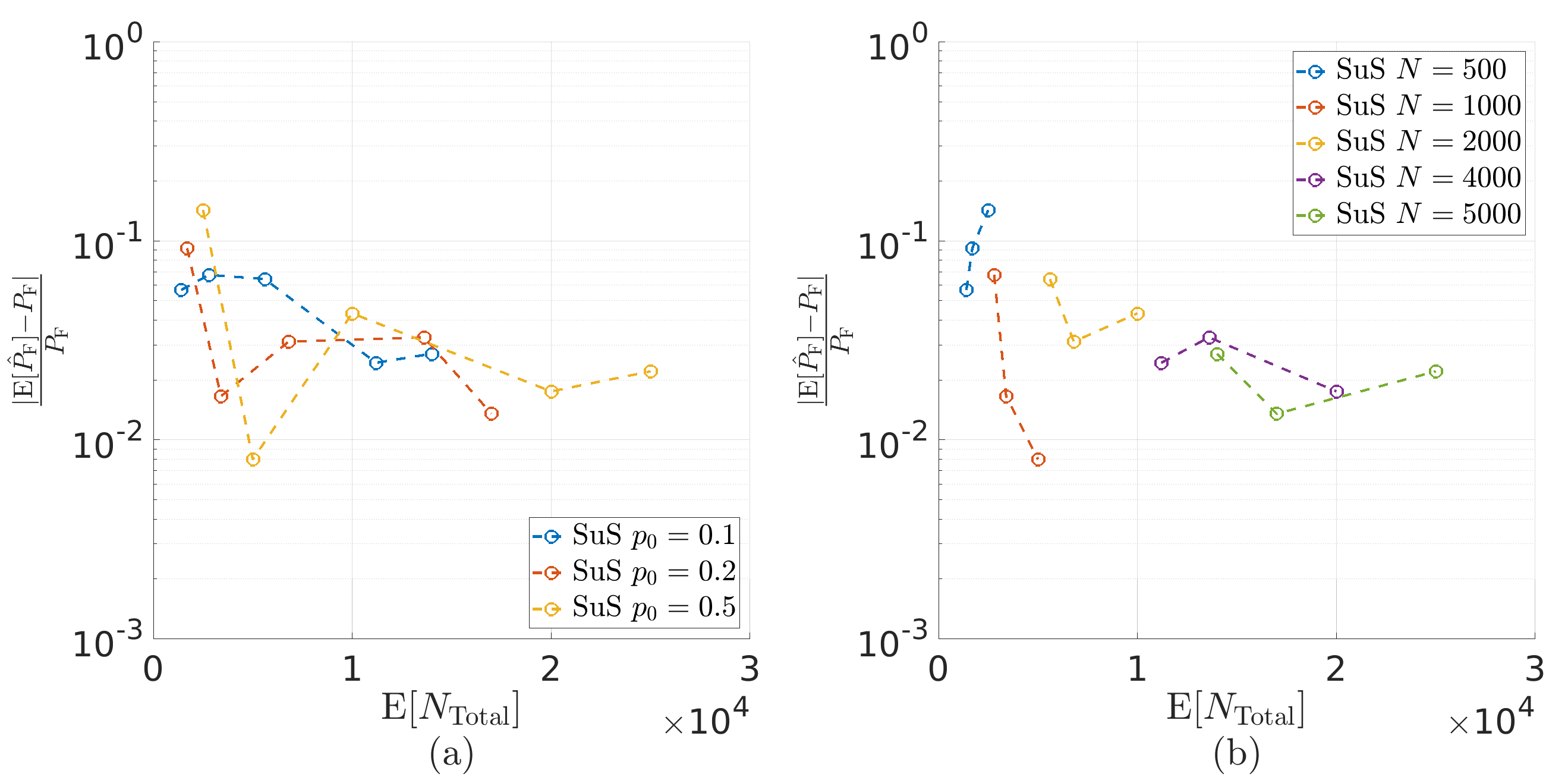}
    \caption{For the four failure branches function, Eq.~\eqref{ffbf}, the relative error of the standard implementation as a function of the average number of evaluations, over $20$ independent runs, for (a) the conditional probabilities $p_0 = (0.1, 0.2, 0.5)$ with the corresponding number of samples $N = (500, 1000, 2000, 4000, 5000)$, and (b) the number of samples $N = (500, 1000, 2000, 4000, 5000)$ with the corresponding conditional probabilities $p_0 = (0.1, 0.2, 0.5)$.}
    \label{g5sus}
\end{figure}

\begin{figure}
    \centering
    \includegraphics[scale=0.3]{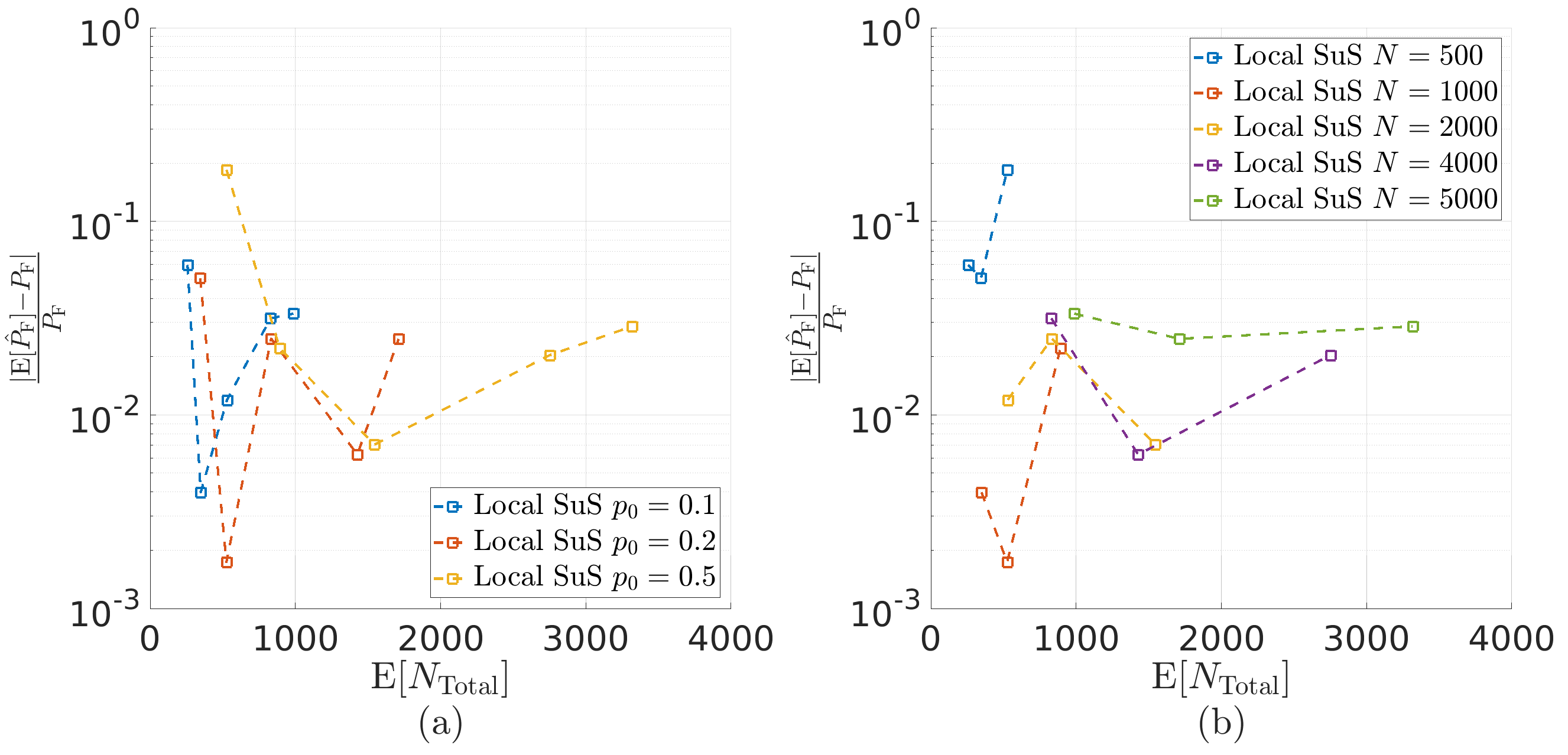}
    \caption{For the four failure branches function, Eq.~\eqref{ffbf}, the relative error of the local subset approach as a function of the average number of evaluations, over $20$ independent runs, for (a) the conditional probabilities $p_0 = (0.1, 0.2, 0.5)$ with the corresponding number of samples $N = (500, 1000, 2000, 4000, 5000)$, and (b) the number of samples $N = (500, 1000, 2000, 4000, 5000)$ with the corresponding conditional probabilities $p_0 = (0.1, 0.2, 0.5)$.}
    \label{g5loc}
\end{figure}

Using the local subset approach, the probability of failure $P_{\rm F}$ is approximated with an average of $348.6$ numerical evaluations, $36\%$ of which are performed at the initial sampling. The relative error $\varepsilon_0$ is less than $4\%$, while the relative error $\varepsilon$ compared to the Monte Carlo estimation is less than $2\%$. Local approximations reduce the computation requirements by $87.5\%$. The SMART algorithm \cite{Bourinet:2011} employs support vector machines in the subset simulation method to approximate the probability of failure under the same conditions as above, with $N_{\rm Total}= 2035$ and the relative error less than $3\%$. The limit-state function, Eq.~\eqref{ffbf}, is used nearly six times more than with the local subset approach. Figures \ref{g5sus} and \ref{g5loc} show the performance of the standard implementation and the local subset approach for different values of $p_0$ and $N$. The local subset approach minimizes the relative error for $p_0 = 0.2$ and $N=1000$. Figure \ref{2g}a illustrates the performance of the local subset approach in the probability space. The exact failure regions are accurately explored and estimated.

\subsection{Example 3 - Hypersphere limit-state function}
Here, the failure region is defined with the samples $\theta_i$ located outside of a hypersphere with radius $\tau$ \cite{betz:2018}

\begin{equation}\label{g3}
    g_3(\theta) = 1 - \frac{\Vert\theta\Vert_2^2}{\tau^2} - \frac{\theta_1}{\tau}\Bigg[\frac{1-(\frac{\Vert\theta\Vert_2}{\tau})^\nu)}{1+(\frac{\Vert\theta\Vert_2}{\tau})^\nu}\Bigg],
\end{equation}
where $\nu \in [0,4]$ modifies the gradient of the limit-state function in $\theta_1$ direction. The failure domain is independent of $\nu$ for this range. The reference probability of failure for $\nu=2$, $d=2$ and $\tau=5.26$ is ${\rm Pr}[g(\theta)\leq0] = 1 \times 10^{-6}$. The exact solution can be derived with the upper and lower incomplete gamma functions \cite{betz:2018}.
\begin{table}[ht]
\normalsize
\begin{adjustbox}{width=\columnwidth,center}
\begin{tabular}{c c c c c c c c c c} 
\hline\hline 
Case & $\mathbb{E}[P_{\rm F}^{\rm MC}]$ & $\mathbb{E}[P_{\rm F}^{\rm SuS}]$ & $\mathbb{E}[P_{\rm F}^{\rm Local}]$ & $\sigma[P_{\rm F}^{\rm Local}]$ & $\varepsilon$ & $\varepsilon_0$ &$\mathbb{E}[\widehat{N}_{0}]$ & $\mathbb{E}[\widehat{N}_{\rm Total}]$ & $\mathbb{E}[N_{\rm Total}]$\\ [0.5ex] 
\hline 
$g_{3}(\theta)$ & 1.0e-6 & 0.8e-6 & 0.88e-6 & 1.4e-6 & 0.12 & 0.10 & 208 & 1103.1 & 6400\\[1ex] 
\hline 
\end{tabular}
\end{adjustbox}
\caption{Local subset approach for the hypersphere limit-state function averaged over 20 independent runs with $p_0=0.1$ and $N = 1000$.} 
\label{table:3} 
\end{table}

\begin{figure}[ht]
    \centering
    \includegraphics[scale=0.3]{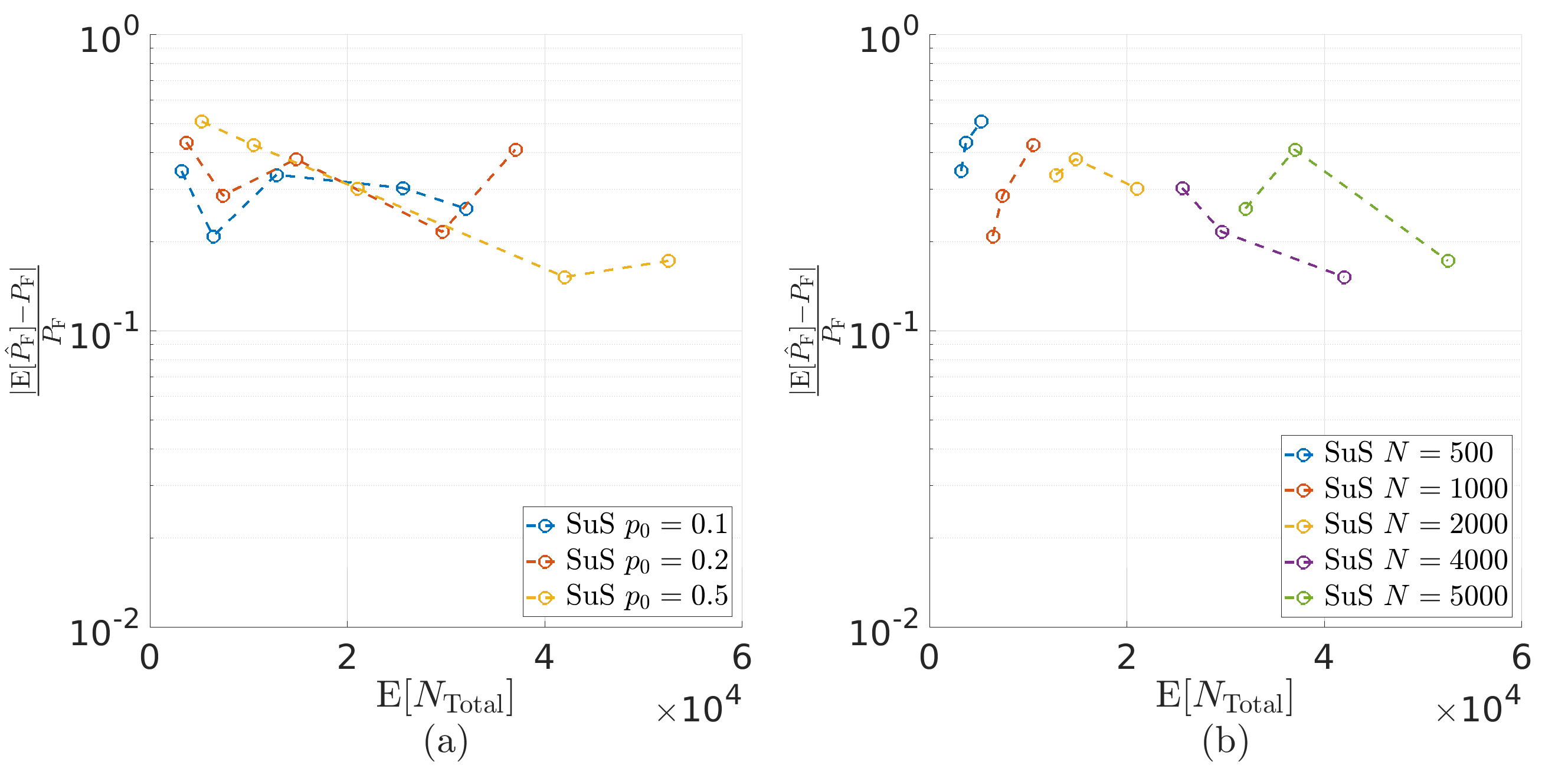}
    \caption{For the hypersphere limit-state function, Eq.~\eqref{g3}, the relative error of the standard implementation as a function of the average number of evaluations, over $20$ independent runs, for (a) the conditional probabilities $p_0 = (0.1, 0.2, 0.5)$ with the corresponding number of samples $N = (500, 1000, 2000, 4000, 5000)$, and (b) the number of samples $N = (500, 1000, 2000, 4000, 5000)$ with the corresponding conditional probabilities $p_0 = (0.1, 0.2, 0.5)$.}
    \label{g9sus}
\end{figure}

\begin{figure}[ht]
    \centering
    \includegraphics[scale=0.3]{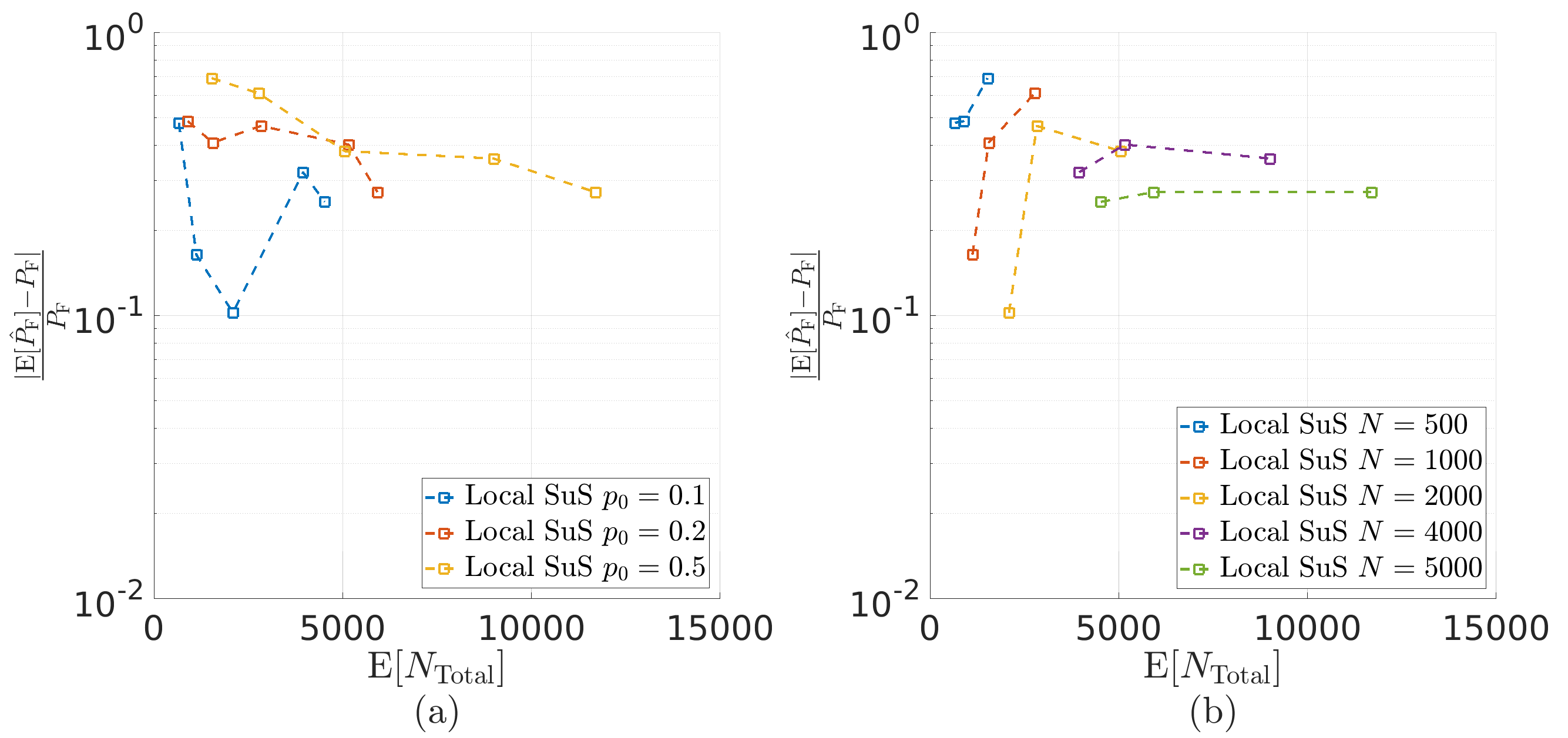}
    \caption{For the hypersphere limit-state function, Eq.~\eqref{g3}, the relative error of the local subset approach as a function of the average number of evaluations, over $20$ independent runs, for (a) the conditional probabilities $p_0 = (0.1, 0.2, 0.5)$ with the corresponding number of samples $N = (500, 1000, 2000, 4000, 5000)$, and (b) the number of samples $N = (500, 1000, 2000, 4000, 5000)$ with the corresponding conditional probabilities $p_0 = (0.1, 0.2, 0.5)$.}
    \label{g9loc}
\end{figure}

For the hypersphere limit-state function, Eq.~\eqref{g3}, with $p_0=0.1$, the number of failure levels is $L-1=6$. This is a practical example to analyze our approach for more levels and smaller failure probabilities. The results exhibit a remarkable performance of the local subset approach. The relative error $\varepsilon_0$ is less than $10\%$ with the reduction in the computational requirements of $82.8\%$, see Table \ref{table:3}. In comparison to the exact solution, the relative error is up to $12\%$. Figure \ref{2g}b shows that the failure thresholds are accurately estimated with the nested condition satisfied. For this example, we design locally a Gaussian process with the constant trend. In Figs.~\ref{g9sus} and \ref{g9loc} we plot the relative error as function of the expected value of the total number of limit-state evaluations for different conditional probabilities $p_0$ and different numbers $N$ of initial limit-state evaluations. The local subset approach requires fewer evaluations than the standard implementation with the adaptive MCMC algorithm. The minimal relative error for the local subset approach is attained at $p_0 = 0.1$ and $N=2000$.

\subsection{Example 4 - Nonlinear oscillator}
We here adapt the nonlinear oscillator from \cite{pls4}, which is a hysteretic oscillator under stochastic loading governed by
\begin{equation}\label{nonos}
    m_0\Ddot{u}(t) + a\dot{u}(t) + a_0[\alpha u(t) + (1-\alpha)u_y z(t)] = \Psi(t),
\end{equation}
where $u(t), \dot{u}(t)$ and $\Ddot{u}(t)$ are the displacement, velocity and acceleration of the oscillator in time $t$. We select the design parameters as $m_0=6\cdot10^4$, $a_0 = 5 \cdot 10^6$, $a =2m_0\zeta \sqrt{a_0/m_0}$, $\zeta = 5 \%$ and $u_y = 0.04$. The parameter $\alpha=0.1$ is introduced to control the degree of hysteresis. The parameter $z(t)$ is governed by the Bouc-Wen hysteresis law \cite{pls4}. The loading $\Psi(t)$ is a seismic load model as a white noise, which is a time-series. It is discretized in the frequency domain as \cite{pls4}
\begin{equation}
    \Psi(t)=-m_0S_w \sum_{n=1}^{d/2} [\theta_n \cos(w_nt) + \theta_{d/2+n} \sin(w_nt)].
\end{equation}
Here $\theta_n$, $n=1,...,d$ are independent standard Gaussian random variables, $w_n = n\Delta w$, $\Delta w = 30\pi/d$, the cut-off frequency is $w_{\rm cut} = 15\pi$ and $S_w = \sqrt{2S_0\Delta w}$ with the intensity of the white noise $S_0=0.03$. We define $d=300$ to approximate the probability of failure ${\rm Pr}[u(8s) + 0.3 \leq 0]$ for the displacement of the oscillator at $t=8s$. The reference probability of failure is estimated to $P_{\rm F}^{\rm MC}= 8.3 \times 10^{-4}$ using the simple Monte Carlo method with $N_{\rm MC} = 1 \times 10^6$.

\begin{table}
\normalsize
\begin{adjustbox}{width=\columnwidth,center}
\begin{tabular}{c c c c c c c} 
\hline\hline 
Case & $\mathbb{E}[P_{\rm F}^{\rm SuS}]$ & $\mathbb{E}[P_{\rm F}^{\rm Local}]$ & $\sigma[P_{\rm F}^{\rm Local}]$ & $\varepsilon_0$ & $\mathbb{E}[\widehat{N}_{\rm Total}]$ & $\mathbb{E}[N_{\rm Total}]$\\ [0.5ex] 
\hline 
(a) $u(8s)+0.3$ & 5.8e-4 & 0.8e-4 & 4.5e-5 & 0.86 & 2420.1 & 3700\\ 
\hline
(b) $u(8s)+0.3$ & 5.9e-4 & 3.3e-4 & 1.2e-4 & 0.44 & 3992.5 & 6000\\ 
\hline 
(c) $u(8s)+0.3$ & 7.2e-4 & 4.4e-4 & 4.9e-5 & 0.38 & 7457.9 & 1.25e4\\ 
\hline 
(d) $u(8s)+0.3$ & 5.8e-4 & 2.8e-4 & 7.0e-5 & 0.51 & 1674.5 & 3700\\ 
\hline 
\end{tabular}
\end{adjustbox}
\caption{Local PLS-Gaussian process approach for the nonlinear oscillator for (a) $d=300$ with $p_0=0.1$ and $N=1000$, (b) $d=300$ with $p_0=0.5$ and $N=1000$, (c) $d=300$ with $p_0=0.1$ and $N=5000$, and (d) $d_{\rm PCA}=110$ with $p_0=0.1$ and $N=1000$.} 
\label{table:4} 
\end{table}

The local PLS-Gaussian process approach estimates the probability of failure $P_{\rm F}$ with the relative error $\varepsilon_0$ less than $86\%$ and with the efficiency of $34.6\%$, see Table \ref{table:4}. By increasing $N=5000$, the relative error drops to $38\%$ with the efficiency above $59\%$ with respect to the standard implementation. In general, the relative errors for the local subset approach using GP repression with PLS are substantial. 

\begin{figure}[ht]
    \centering
    \includegraphics[scale=0.38]{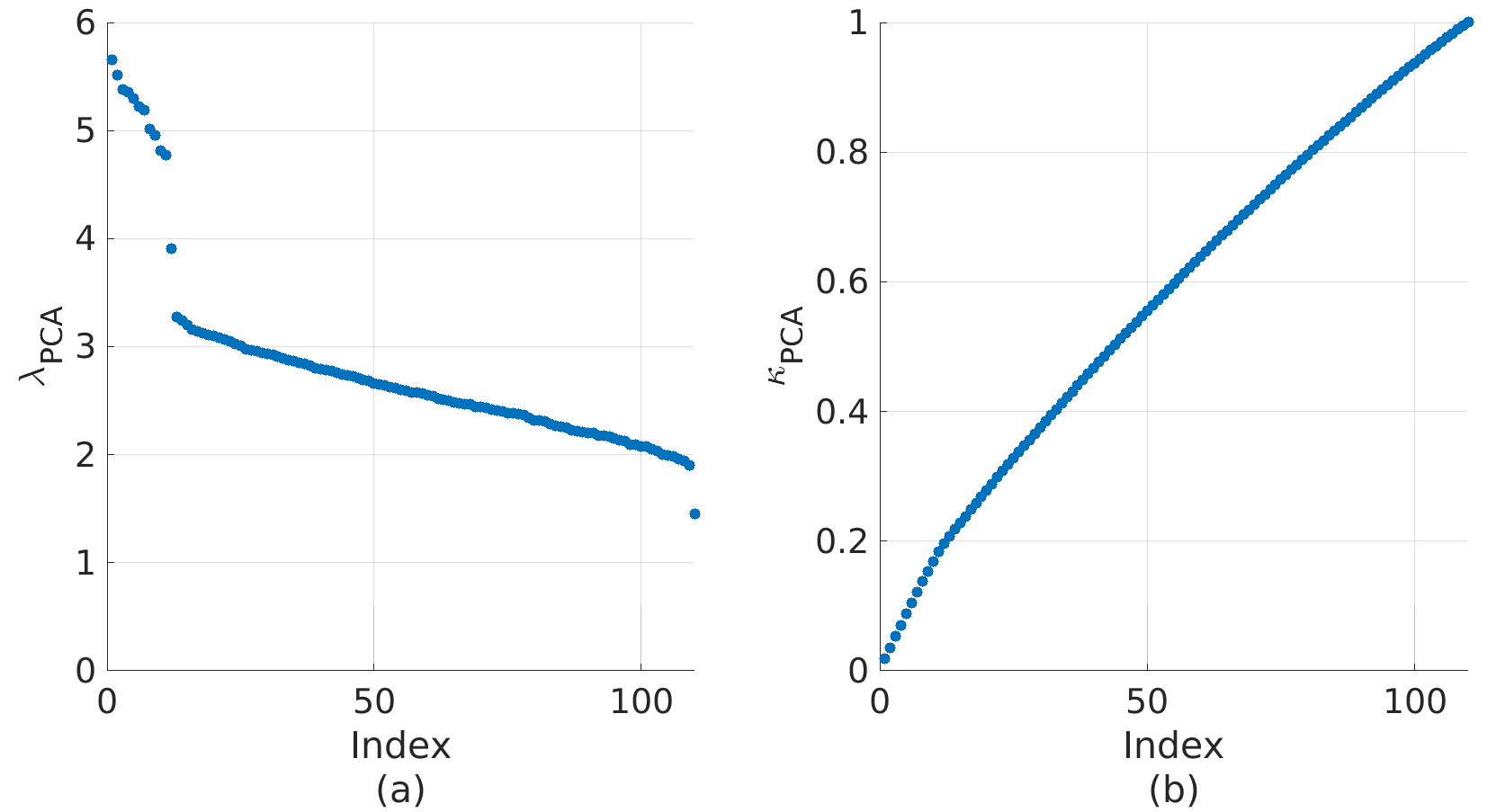}
    \caption{(a) PCA eigenvalues for the loading, and (b) Total variation using the PCA eigenvalues.}
    \label{pca}
\end{figure}

The input parameters $\theta$ drawn from the standard normal distribution are the Fourier coefficients of the loading $\Psi(t)$, which is a time-series that resembles white noise. Employing the principal component analysis (PCA) on the input parameters $\theta$ would be pointless because $\theta$ are iid variables. However, $\Psi(t)$ as a time-series can contain a sufficient low-dimensional subspace in contrast to the input parameters. The loading $\Psi(t)$ at each time $t$ can be used as the input parameter instead of the Fourier coefficients. Therefore, we initially generate 5000 different independent realizations of $\Psi(t)$ using 300 Fourier coefficients. The loading $\Psi(t)$ is discretized with 110 time steps for $t=0,\dots,8$. The realizations are used to estimate the eigenvalues and eigenvectors for the loading $\Psi(t)$, see Fig.~\ref{pca}a. In Fig.~\ref{pca}b, the total variation estimation show that all $110$ elements of $\Psi(t)$ are important to maintain the variation $100\%$. The total variation increases linearly with the discretized elements of $\Psi(t)$. Hence, the eigenvectors are used to project the elements of $\Psi(t)$ to linearly uncorrelated variables. For each projected element of $\Psi(t)$, we define the normal distribution using $5000$ different independent realizations. The projected elements are now used to govern the nonlinear oscillator instead of the Fourier coefficients. This represents the dimension reduction of $63\%$. Thus, using the inverse of the eigenvectors with an independent, uncorrelated realization of the projected elements, we estimate the original loading $\Psi(t)$ for Eq.~\eqref{nonos}. PCA increases the efficiency for the local subset approach to $54.8\%$ for $N=1000$ and $p_0=0.1$, see Table \ref{table:4}. The relative error drops from $90\%$ with respect to the simple Monte Carlo estimation to $66\%$. In comparison with the standard implementation, the relative error drops to $51\%$. A smaller input dimension requires a smaller design set, which eventually generates stable predictions. As we can observe, the relative errors are substantial due to the strong nonlinearity and the dimension of the system. However, the order of the probability of failure of $10^{-4}$ is accurately estimated in all cases.

\section{Conclusion}
We propose a novel approach that uses local surrogates to reduce the cost of the Bayesian approximation in the subset simulation method. Here, we employ Gaussian process regression for each Markov chain proposal to utilize the local regularity of the limit-state function. The posterior variance and the random indicator are used to control errors in predictions. When one of the error indicators is triggered, the refinement procedure employs the limit-state function adequately to improve the prediction or the sample set locally. We use the $U$-function to include a failure threshold in the refinement procedure. The numerical experiments indicate a clear advantage of our local subset approach over the standard implementation of the subset simulation method. The total number of evaluations is reduced by over $80\%$ while maintaining the relative error up to $12\%$. For higher dimensions, the performance is comparable to the standard implementation, but with intensive computations.

To address this, the partial least square (PLS) regression is implemented in the local subset algorithm to define a low-dimensional subspace for a Markov chain proposal. The approach maximizes the squared covariance between the low-dimensional projection of the input parameters and limit-state evaluations. PLS is suitable for expensive numerical models as it does not require gradient evaluations and provides an adequate reduction even for limited sample sets. However, the efficiency of the local subset approach decreases to $34\%$ with significant relative errors. Nevertheless, the order for the probability of failure is accurately estimated in most cases.

Our algorithms can still be improved, especially for high-dimensional numerical experiments. Expensive forward models typically have adjoint solvers to estimate gradients efficiently. Therefore, we plan to examine the possibility of including gradients by using the active-subspace analysis for local predictions.

\section*{Acknowledgements}
The authors would like to thank Youssef M. Marzouk for productive discussions, useful comments, and suggestions. This research was funded by the DeRisk project,  Innovation  Fund  Denmark,  grant number 4106-00038B. K\v S especially acknowledges the support from Otto M\o nsteds Fond, Danish Agency for Science and Higher Education, and Massachusetts Institute of Technology (MIT) during his research stay at MIT

\bibliographystyle{unsrt}  
\bibliography{references}  

\begin{thebibliography}{10}

\bibitem{rose}
J.~E. Hurtado.
\newblock {\em Structural Reliability - Statistical Learning Perspectives}.
\newblock Springer-Verlag Berlin Heidelberg, 1 edition, 2004.

\bibitem{nataf}
R.~Lebrun and A.~Dutfoy.
\newblock An innovating analysis of the nataf transformation from the copula
  view point.
\newblock {\em Probabilistic Engineering Mechanics}, 24(3):312--320, 2009.

\bibitem{Iason:2015}
I.~Papaioannou, W.~Betz, K.~Zwirglmaier, and D.~Straub.
\newblock {MCMC} algorithms for {S}ubset {S}imulation.
\newblock {\em Probabilistic Engineering Mechanics}, 41:89--103, 2015.

\bibitem{foam1}
R.~Rackwitz.
\newblock Reliability analysis - a review and some perspectives.
\newblock {\em Structural Safety}, 23(4):365--395, 2001.

\bibitem{foam2}
M.A. Valdebenito, H.J. Pradlwarter, and G.I. Schu\"eller.
\newblock The role of the design point for calculating failure probabilities in
  view of dimensionality and structural nonlinearities.
\newblock {\em Structural Safety}, 32(2):101--111, 2010.

\bibitem{mcbook}
A.~B. Owen.
\newblock {\em Monte Carlo theory, methods and examples}.
\newblock Open Access, 2013.

\bibitem{mcvar}
C.~Bucher.
\newblock Adaptive sampling $-$ an iterative fast {M}onte {C}arlo procedure.
\newblock {\em Structural Safety}, 5(2):119--126, 1988.

\bibitem{mcvar1}
PT. de~Boer, D.~P. Kroese, S.~Mannor, and R.~Y. Rubinstein.
\newblock A tutorial on the cross-entropy method.
\newblock {\em Annals of Operations Research}, 134(1):19--67, 2005.

\bibitem{zuev:2012}
K.~M. Zuev, J.~L. Beck, S.~K. Au, and L.~S. Katafygiotis.
\newblock Bayesian post-processor and other enhancements of {S}ubset
  {S}imulation for estimating failure probabilities in high dimensions.
\newblock {\em Computers and Structures}, 92-93:283--296, 2012.

\bibitem{au:2001}
S.~K. Au and J.~L. Beck.
\newblock Estimation of small failure probabilities in high dimensions by
  subset simulation.
\newblock {\em Probabilistic Engineering Mechanics}, 16(4):263--277, 2001.

\bibitem{hmc}
Z.~Wang, M.~Broccardo, and J.~Song.
\newblock Hamiltonian {M}onte {C}arlo methods for subset simulation in
  reliability analysis.
\newblock {\em Structural Safety}, 76:51--67, 2019.

\bibitem{Conrad:2016}
P.~R. Conrad, Y.~M. Marzouk, N.~S. Pillai, and A.~Smith.
\newblock {A}ccelerating {A}symptotically {E}xact {MCMC} for {C}omputationally
  {I}ntensive {M}odels via {L}ocal {A}pproximations.
\newblock {\em Journal of the American Statistical Association},
  111:516:1591--1607, 2016.

\bibitem{Bourinet:2011}
J.~M. Bourinet, F.~Deheeger, and M.~Lemaire.
\newblock Assessing small failure probabilities by combined subset simulation
  and {S}upport {V}ector {M}achines.
\newblock {\em Structural Safety}, 32(6):343--353, 2011.

\bibitem{Bourinet:2016}
J.~M. Bourinet.
\newblock Rare-event probability estimation with adaptive support vector
  regression surrogates.
\newblock {\em Reliability Engineering and System Safety}, 150(2016):210--221,
  2016.

\bibitem{beck1}
J.~Bect, L.~Ling, and E.~Vazquez.
\newblock Bayesian subset simulation.
\newblock {\em SIAM/ASA Journal on Uncertainty Quantification}, 5(1):762--786,
  2017.

\bibitem{mlmc}
E.~Ullmann and I.~Papaioannou.
\newblock Multilevel estimation of rare events.
\newblock {\em SIAM/ASA Journal on Uncertainty Quantification}, 3(1):922--953,
  2015.

\bibitem{pls2}
G.~Shen, M.~Lesnoff, V.~Baeten, P.~Dardenne, F.~Davrieux, H.~Ceballos,
  J.~Belalcazar, D.~Dufour, Z.~Yang, L.~Han, and J.~A. Fernández~Pierna.
\newblock Local partial least squares based on global {PLS} scores.
\newblock {\em Journal of Chemometrics}, 33(5):e3117, 2019.

\bibitem{iason46}
L.~Tierney.
\newblock Markov chains for exploring posterior distributions.
\newblock {\em The Annals of Statistics}, 22(4):1701--1762, 1994.

\bibitem{cerou:2012}
F.~C\'erou, P.~Del~Moral, T.~Furon, and A.~Guyader.
\newblock Sequential {M}onte {C}arlo for rare event estimation.
\newblock {\em Statistics and Computing}, 22:795--808, 2012.

\bibitem{susprob}
K.~Breitung.
\newblock The geometry of limit state function graphs and subset simulation:
  Counterexamples.
\newblock {\em Reliability Engineering \& System Safety}, 182:98--106, 2019.

\bibitem{gp3}
A.~V. Vecchia.
\newblock Estimation and model identification for continuous spatial processes.
\newblock {\em Journal of the Royal Statistical Society: Series B
  (Methodological)}, 50(2):297--312, 1988.

\bibitem{gp2}
M.~L. Stein, Z.~Chi, and L.~J. Welty.
\newblock Approximating likelihoods for large spatial data sets.
\newblock {\em Journal of the Royal Statistical Society: Series B (Statistical
  Methodology)}, 66(2):275--296, 2004.

\bibitem{gp1}
E.~Snelson and Z.~Ghahramani.
\newblock Local and global sparse {G}aussian process approximations.
\newblock In {\em Proceedings of the Eleventh International Conference on
  Artificial Intelligence and Statistics (AISTATS-07)}, pages 524--531, 2007.

\bibitem{gramacy}
R.~B. Gramacy and D.~W. Apley.
\newblock Local {G}aussian process approximation for large computer
  experiments.
\newblock {\em Journal of Computational and Graphical Statistics}, 24:561--578,
  2015.

\bibitem{bruno:2016}
R.~Sch\"obi, B.~Sudret, and S.~Marelli.
\newblock Rare event estimation using {P}olynomial-{C}haos {K}riging.
\newblock {\em ASCE-ASME Journal of Risk and Uncertainty in Engineering
  Systems, Part A: Civ. Eng.}, 3(2), 2017.

\bibitem{davis:2018}
A.~D. Davis.
\newblock {\em {P}rediction under uncertainty: from models for
  marine-terminating glaciers to {B}ayesian computation}.
\newblock PhD thesis, Massachusetts Institute of Technology, 2018.

\bibitem{li:2010}
J.~Li and D.~Xiu.
\newblock Evaluation of failure probability via surrogate models.
\newblock {\em Journal of Computational Physics}, 229(23):8966--8980, 2010.

\bibitem{pls3}
K.~Hazama and M.~Kano.
\newblock Covariance-based locally weighted partial least squares for
  high-performance adaptive modeling.
\newblock {\em Chemometrics and Intelligent Laboratory Systems},
  146(2015):55--62, 2015.

\bibitem{pls1}
M.A. Bouhlel, N.~Bartoli, A.~Otsmane, and J.~Morlier.
\newblock Improving kriging surrogates of high-dimensional design models by
  partial least squares dimension reduction.
\newblock {\em Structural and Multidisciplinary Optimization}, 53:935--952,
  2016.

\bibitem{pls4}
I.~Papaioannou, M.~Ehre, and D.~Straub.
\newblock {PLS}-based adaptation for efficient {PCE} representation in high
  dimensions.
\newblock {\em Journal of Computational Physics}, 387:186--204, 2019.

\bibitem{pls5}
K.~Song, T.~Tong, F.~Wu, and Z.~Zhang.
\newblock A novel partial least squares weighting {G}aussian process algorithm
  and its application to near infrared spectroscopy data mining problems.
\newblock {\em Analytical Methods}, 4:1395--1400, 2012.

\bibitem{eralink}
ERA.
\newblock {Engineering Risk Analysis Group TU M\"unich}.
\newblock
  \url{https://www.bgu.tum.de/era/software/software00/subset-simulation/}, Dec.
  2019.

\bibitem{betz:2018}
W.~Betz.
\newblock {\em {B}ayesian inference of engineering models}.
\newblock PhD thesis, Technische Universit\"at M\"unchen, 2018.

\end{thebibliography}

\end{document}